\documentclass[10pt]{aa}  
\usepackage{graphicx}
\usepackage{txfonts}
\begin{document}
   \title{Self-consistent models of quasi-relaxed rotating stellar systems}


   \author{A. L. Varri
          \and
          G. Bertin
          }

   \institute{Dipartimento di Fisica, Universit\`a degli Studi di Milano,
              Via Celoria 16, I-20133 Milano, Italy\\
              \email{anna.varri@unimi.it; giuseppe.bertin@unimi.it}}

   \date{Received 19 October 2011; to appear}

 
  \abstract
   {}  
{Two new families of self-consistent axisymmetric truncated equilibrium models for the description of quasi-relaxed rotating stellar systems are presented. The first extends the well-known spherical King models to the case of solid-body rotation. The second is characterized by differential rotation, designed to be rigid in the central regions and to vanish in the outer parts, where the imposed energy truncation becomes effective.}
{The models are constructed by solving the relevant nonlinear Poisson equation for the self-consistent mean-field potential. For rigidly rotating configurations, the solutions are obtained by an asymptotic expansion based on the rotation strength parameter, following a procedure developed earlier by us for the case of tidally generated triaxial models. The differentially rotating models are constructed by means of a spectral iterative approach, with a numerical scheme based on a Legendre series expansion of the density and the potential.}
{The two classes of models exhibit complementary properties. The rigidly rotating configurations are flattened toward the equatorial plane, with deviations from spherical symmetry that increase with the distance from the center. For models of the second family, the deviations from spherical symmetry are strongest in the central region, whereas the outer parts tend to be quasi-spherical. The relevant parameter spaces are thoroughly explored and the corresponding intrinsic and projected structural properties are described. Special attention is given to the effect of different options for the truncation of the distribution function in phase space.}
{Models in the moderate rotation regime are best suited to applications to globular clusters. For general interest in stellar dynamics, at high values of the rotation strength the differentially rotating models tend to exhibit a toroidal core embedded in an otherwise quasi-spherical configuration. Physically simple analytical models of the kind presented here provide insights into dynamical mechanisms and may be a useful basis for more realistic investigations carried out with the help of N-body simulations.}


   \maketitle

\section{Introduction}

A large class of stellar systems is expected to be in a quasi-equilibrium state characterized by a distribution function not far from a Maxwellian. Mechanisms that are thought to operate in this direction range from standard two-body gravitational scattering (Chandrasekhar \cite{Cha43}) to violent relaxation, instabilities, and phase mixing for collisionless systems (Lynden-Bell \cite{Lyn67}). In particular, for relatively small stellar systems, such as globular clusters, two-star collisional relaxation is thought to be responsible for such quasi-relaxed condition, while for large stellar systems, such as elliptical galaxies, incomplete violent relaxation has likely acted during the formation process so as to bring the system to a state in which significant (radially biased) anisotropy is present only for weakly bound stars. To be sure, full relaxation would lead to a singular state, characterized by infinite mass; the so-called isothermal sphere solution may provide an approximate representation only of the inner parts of real stellar systems.

In line with the above considerations, several physically motivated self-consistent finite-mass models (thus characterized by significant, but not arbitrary, deviations from a pure Maxwellian) have been constructed and studied, with interesting astronomical applications. For the description of partially relaxed systems, self-consistent anisotropic models have been introduced to describe the products of incomplete violent relaxation (see Bertin \& Stiavelli \cite{BerSti93} and Trenti et al. \cite{Tre05}; and references therein) . For the description of quasi-relaxed globular clusters, the spherical isotropic King (\cite{Kin66}) models, which incorporate in a heuristic way the concept of tidal truncation, provide a standard reference framework (see Zocchi et al. \cite{ZBV11} for a photometric and kinematic study of a sample of Galactic globular clusters under different relaxation conditions). Recently, King models have been extended to the full triaxial case, to include explicitly the three-dimensional role of an external tidal field in the simple case of a circular orbit of the cluster within the host galaxy (Bertin \& Varri \cite{BerVar08}, Varri \& Bertin \cite{VarBer09}, in the following denoted as Paper I and II, respectively; see also Heggie \& Ramamani \cite{HegRam95}); these triaxial models can thus be seen as a collisionless analogues of the purely tidal Roche ellipsoids.

For astronomical applications, one of the main drivers (or empirical clues) in the above modeling investigations is the issue of the origin of the observed geometry of stellar systems, being it known that pressure anisotropy, tides, or rotation can be responsible, separately, for deviations from spherical symmetry.  In the above-mentioned studies, very little attention was actually placed on the role of rotation. For ellipticals, most of the attention that led to the development of stellar dynamical models, after the first kinematical measurements became available in the mid-70s, was taken by the study of the 
curious behavior of pressure-supported systems in the presence of anisotropic orbits  (see also Schwarzschild \cite{Sch79}, \cite{Sch82}; de Zeeuw \cite{Zee85}). In contrast, very little effort has been made in modeling rotation-dominated ellipticals, even though the entire low-mass end of the distribution of elliptical galaxies might be consistent with a picture of rotation-induced flattening (e.g., see Davies et al. \cite{Dav83}, and more recently Emsellem et al. \cite{Ems11}); similar comments apply to bulges.

For globular clusters, given the fact that they only exhibit modest amounts of flattening and given the success of the spherical King models, little work has been carried out in the direction of stationary self-consistent rotating models (with some notable exceptions, that is  Wolley \& Dickens \cite{WooDic62}, Lynden-Bell \cite{Lyn62}, Kormendy \& Anand \cite{KorAna71}, Lupton \& Gunn \cite{LupGun87}, and Lagoute \& Longaretti \cite{LagLon96}). Therefore, as far as rotation-dominated systems are concerned, much of the currently available modeling tools go back to the pioneering work of Prendergast \& Tomer (\cite{PreTom70}), Wilson (\cite{Wil75}), and Toomre (\cite{Too82}), intended to describe ellipticals, and of Jarvis \& Freeman (\cite{JarFre85}) and Rowley (\cite{Row88}), devoted to bulges. In general, we may say that only very few rotating models with explicit distribution function are presently known (for a recent example, see Monari et al. in preparation). In this context, one should also mention the interesting work by Vandervoort (\cite{Van80}) on the collisionless analogues of the Maclaurin and Jacobi ellipsoids.

On the empirical side, a deeper study of quasi-relaxed rotating stellar systems is actually encouraged by the widespread conviction that the geometry of 
the inner parts of globular clusters, where some flattening is noted, should not be the result of tides, but rather of rotation (King \cite{Kin61}). 
In other words, it is frequently believed that tides and pressure anisotropy (and dust obscuration), even though playing some role in individual cases, should not  be considered as the primary explanation of the observed flattening of Galactic globular clusters. Such conclusion is suggested by the White \& Shawl (\cite{WhiSha87}) database of ellipticities. However, for this database we note that: (i) the cluster flattening values do not all refer to a standard isophote, such as the cluster half-light radius (as also noted by van den Bergh \cite{Ber08}), (ii) the data mostly refer to the inner regions. Such limitations are crucial because there is observational evidence that the ellipticity of a cluster depends on radius (see Geyer et al. \cite{Gey83}). In turn, recent studies by Chen \& Chen (\cite{CheChe10}), in which these restrictions are addressed more carefully, suggest that Galactic globular clusters are less round than previously reported, especially  those in the region of the Galactic Bulge; at variance with previous analyses, it seems that their major axes preferentially point toward the Galactic Center, as naturally expected if their shape is of tidal origin. 

The connection between flattening and internal rotation has been discussed in detail by means of nonspherical dynamical models in just a handful of cases, in particular for $\omega$ Cen (for an oblate rotator nonparametric model, see Merritt \cite{Mer97}; for an orbit-based analysis, see van de Ven \cite{Ven06}; for an application of the Wilson \cite{Wil75} models, see Sollima et al. \cite{Sol09}), 47 Tuc (Meylan \& Mayor \cite{MeyMay86}), M15 (van den Bosch et al. \cite{Bos06}), and M13 (Lupton et al. \cite{Lup87}). In addition, specifically designed 2D Fokker-Planck models (Fiestas et al. \cite{Fie06}) have been applied to the study of M5, NGC 2808, and NGC 5286. We recall that the detection of internal rotation in globular clusters is a challenging task, because the typical value of the ratio of mean velocity to velocity dispersion is only of a few tenths, for example $V/\sigma \approx 0.26, 0.32$ for 47 Tuc and $\omega$ Cen, respectively (Meylan and Mayor \cite{MeyMay86}; for a summary of the results for several Galactic objects, see also Table 7.2 in Meylan \& Heggie \cite{MeyHeg97}). However, great progress made in the acquisition of photometric and kinematical information, and in particular of the proper motion of thousands of stars (for  $\omega$ Cen, see van Leeuwen et al. \cite{Lee00}, Anderson \& van der Marel \cite{AndvdM2010};  for 47 Tucanae, see Anderson \& King \cite{AndKin03} and McLaughlin et al. \cite{Lau06}), makes this goal within reach (see Lane et al. \cite{Lan09}, Lane et al. \cite{Lan10} for new kinematical measurements, in which rotation, when present, is clearly identified).  

On the theoretical side, two general questions provide further motivation to study quasi-relaxed rotating stellar systems. On the one hand, many papers have studied the role of rotation in the general context of the dynamical evolution of globular clusters, but a solid interpretation is still missing. Early investigations (Agekian \cite{Age58}, Shapiro \& Marchant \cite{ShaMar76}) suggested that initially rotating systems should experience a loss of angular momentum induced by evaporation, that is, angular momentum would be removed by stars escaping from the cluster. Because of the small number of particles, N-body simulations were initially (Aarseth \cite{Aar69}, Wielen \cite{Wie74}, and Akiyama \& Sugimoto \cite{AkiSug89}) unable to clearly describe the complex interplay between relaxation and rotation. Later investigations, primarily based on a Fokker-Planck approach (Goodman \cite{Goo83}, Einsel \& Spurzem \cite{EinSpu99}, Kim et al. 
\cite{Kim02}, and Fiestas et al. \cite{Fie06}) have clarified this point, not only by testing the proposed mechanism of angular momentum removal by escaping stars, but also by showing that rotation accelerates the entire dynamical evolution of the system. More recent N-body simulations (Boily \cite{Boi00}, Ernst et al. \cite{Ern07}, and Kim et al. \cite{Kim08}) confirm these conclusions and show that, when a three-dimensional tidal field is included, such acceleration is enhanced even further. The mechanism of angular momentum removal is generally considered to be the reason why Galactic globular clusters are much rounder than the (younger) clusters in the Magellanic Clouds, for which an age-ellipticity relation has been noted (Frenk \& Fall \cite{FreFal82}), but other mechanisms might operate to  produce the observed correlations (Meylan \& Heggie \cite{MeyHeg97}, van den Bergh \cite{Ber08}).

On the other hand, the role of angular momentum during the initial stages of cluster formation should be better clarified. In the context of dissipationless collapse, relatively few investigations have considered the role of angular momentum in numerical experiments of violent relaxation (e.g., the pioneering studies by Gott \cite{Got73}; see also Aguilar \& Merritt \cite{AguMer90}). Interestingly, the final equilibrium configurations resulting from such collisionless collapse show a central region with solid body rotation, while the external parts are characterized by differential rotation.    

In conclusion, mastering the internal structure of spheroidal and triaxial stellar systems through a full spectrum of models, including rotation, is 
a prerequisite for studies of many empirical and theoretical issues. In addition, it is required for the interpretation of the relevant scaling laws (such as the Fundamental Plane, which appears to extend from the brightest, pressure supported ellipticals down to the low-luminosity end of the distribution of early-type galaxies, and possibly further down to the domain of globular clusters) and for investigations aimed at identifying the presence of invisible matter (in the form of central massive black holes or diffuse dark matter halos) from stellar dynamical measurements. It would thus be desirable to construct rotating models to be tested on low luminosity ellipticals, bulges, and globular clusters, especially now that important progress has been made in the collection and analysis of kinematical data. Presumably, many of these stellar systems are quasi-relaxed. In which directions should we explore deviations from the strictly relaxed case?

In the present paper, we consider two families of axisymmetric rotating models: the first one is characterized by the presence of solid-body rotation and isotropy in velocity space (see Appendix B in Paper I for a brief introduction). Indeed, full relaxation in the presence of nonvanishing total angular momentum suggests the establishment of solid-body rotation through the dependence of the relevant distribution function $f=f(H)$ on the Jacobi integral $H = E - \omega J_z $ (see Landau \& Lifchitz \cite{LanLif67}, p. 125). But, for applications to real stellar systems, one may take advantage of the fact that the collisional relaxation time may be large in the outer regions, so that in the outer parts the constraint of solid-body rotation might be released. In particular, for globular clusters we may argue that the outer parts fall into a tide-dominated regime, for which evaporation tends to erase systematic rotation even if initially present, as confirmed by the above-mentioned study based on the Fokker-Planck method. For the truncation, we may then consider a heuristic prescription to simulate the effects of tides, much like for the spherical King models. 

In view of possible applications to globular clusters, we thus consider a second class of axisymmetric rotating models based on a distribution 
function dependent only on the energy and on the z-component of the angular momentum $f=f(I)$ where $I=I(E,J_z)$, with the property that $I \sim E$ 
for stars with relatively high $z$-component of the angular momentum, while $I \sim H = E - \omega J_z$ for relatively low values of $J_z$. Such models are indeed defined in order to have differential rotation, designed to be rigid in the center and to vanish in the outer parts, where the energy truncation becomes effective. As far as the velocity dispersion is concerned, this family may show a variety of profiles (depending on the values of the relevant free parameters), all of them characterized by the presence of isotropy in the central region.  We thus add two classes of self-consistent models to the relatively short list of rotating stellar dynamical models currently available.

One aspect that plays an important role in defining a physically motivated distribution function, which often goes unnoticed (but see Hunter \cite{Hun77}, 
Davoust \cite{Dav77}, and Rowley \cite{Row88}), is the choice of the truncation prescription in phase space. The advantages and the limitations of alternative options available for the second family of models will be discussed in detail. In this context, we will also address the issue of whether these differentially rotating models fall within the class of systems for which rotation is constant on cylinders.

The paper is structured as follows. The properties of the family of rigidly rotating models, constructed on the basis of general statistical mechanics considerations, are illustrated in Sect.~\ref{Unif}. The family of differentially rotating models, designed for the application to rotating globular clusters, is introduced in Sect.~\ref{Diff}, where we briefly describe the method used for the solution of the self-consistent problem and discuss the relevant parameter space. Section \ref{DiffIntr} is devoted to a study of the intrinsic properties and Sect.~\ref{DiffProj} to the projected observables derived from differentially rotating models. 
After illustrating in Sect.~\ref{Trun} the effect of different truncation prescriptions in phase space, we summarize the results and present our conclusions in Sect.~\ref{Concl}. The appendices are devoted, respectively, to a discussion of the nonrotating limit of our families of models, to the details of the alternative truncation option for the second family, and to a description of the code used for the construction of the differentially rotating configurations. A study of the dynamical stability and of the long-term evolution of the families of models introduced here will be addressed by means of an extensive survey of specifically designed N-body simulations and will be presented in following papers (Varri et al. \cite{VarAAS}; Varri et al., in preparation).

\section{Rigidly rotating models}
\label{Unif}
\subsection{The distribution function}
\label{Unif1}

The construction of rigidly rotating configurations characterized by nonuniform density is a classical problem in the theory of rotating stars, starting with Milne (\cite{Mil23}) and Chandrasekhar (\cite{Cha33}), but it basically remained limited to the study of a fluid with polytropic equation of state, for which the solution of the relevant Poisson equation can be obtained by means of a semi-analytical approach (for a comprehensive description, see Chaps.~5 and 10 in Tassoul \cite{Tas78}; for an enlightening presentation of the general problem of rotating compressible masses, see Chap.~9 in Jeans \cite{Jea29}). The reader is referred to Paper I for a discussion of the application of some of the mathematical methods developed in that context to the construction of nonspherical truncated self-consistent stellar dynamical models. The deviations from spherical symmetry studied in Paper I are induced by the presence of a stationary perturbation characterized by a quadrupolar structure, that is, either an external tidal field or internal solid-body rotation. In particular, in Appendix B of Paper I we briefly outlined the extension of the King (\cite{Kin66}) models to the case of internal solid-body rotation, of which, after a short summary of the relevant definitions, we now provide a full description in terms of the relevant intrinsic and projected properties.

\begin{figure}[t]
\hspace{0.5cm}  \includegraphics[height=.3\textheight]{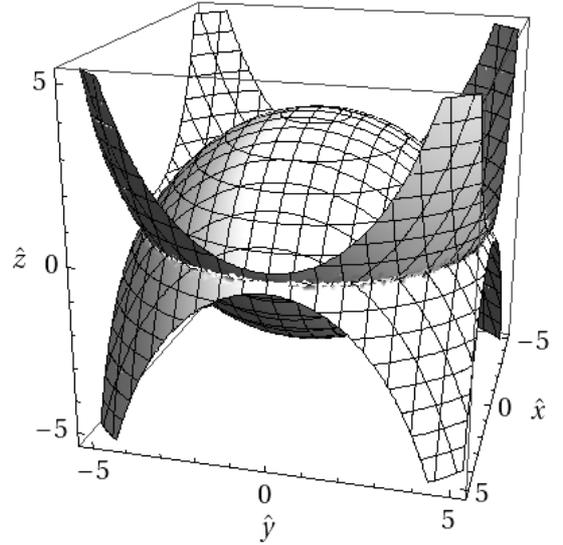}
  \caption{ \normalsize The boundary surface, defined implicitly by $\psi(\hat{{\bf r}})=0$, of a  critical second-order rigidly rotating model with $\Psi=2$. The configuration is axisymmetric; the points on the equatorial plane are saddle points (see Sect. 2.2 for details). Rotation takes place around the $\hat{z}$-axis. The spatial coordinates are expressed in appropriate dimensionless units (see Eq.~(\ref{r0})).}
  \label{crit}
\end{figure}

The extension of the family of King models is performed by considering the distribution function:
\begin{equation}\label{fK}
f_K^r(H)= A\mathrm{e}^{-aH_0}\left[\mathrm{e}^{-a(H-H_0)}-1\right]
\end{equation} 
for $H \le H_0$ and $f_K^r(H)=0$ otherwise, where
\begin{equation}\label{H}
 H = E - \omega J_z 
\end{equation}  
denotes the Jacobi integral, with $\omega$ the angular velocity of the rigid rotation (hence the superscript $r$), assumed to take place around the $z-$axis. The quantities $E$ and $J_z$ are the specific one-star energy and $z$-component of the angular momentum, $H_0$ represents a cut-off constant of the Jacobi integral, while $A$ and $a$ are positive constants. In the corotating frame of reference the Jacobi integral can be written as the sum of the kinetic energy, the centrifugal potential $\Phi_{cen}(x,y)=-(x^2+y^2)\omega^2/2$ (with the equatorial plane $(x,y)$ perpendicular to the rotation axis given by the $z-$axis), and the cluster mean-field potential $\Phi_C$, to be determined self-consistently. Therefore, the dimensionless escape energy can be expressed as:
\begin{equation}
\psi({\bf r})=a\{H_0-[\Phi_C({\bf r})+\Phi_{cen}(x,y)]\}~,
\end{equation}
and the boundary of the cluster, implicitly defined by $\psi({\bf r})=0$, is an equipotential surface for the total potential $\Phi_C+\Phi_{cen}$. Note that, by construction, in the limit of vanishing internal rotation, this family of models reduces to the family of spherical King (\cite{Kin66}) models (see Appendix \ref{AppA} for a summary of the main properties of the family in the nonrotating limit).   

The construction of the models requires the integration of the associated nonlinear Poisson equation, which, after scaling the spatial coordinates with respect to the scale length 
\begin{equation}\label{r0}
r_0= \left(\frac{9}{4\pi G \rho_0 a}\right)^{1/2}~,
\end{equation} 
can be written as
\begin{equation}\label{Poisson}
\hat{\nabla}^2\psi^{(int)}=-9\left[\frac{\hat{\rho}_K(\psi^{(int)})}{\hat{\rho}_K(\Psi)}-2\chi
\right]~,
\end{equation}
where 
\begin{equation}\label{chi}
\chi \equiv \frac{\omega^2}{4 \pi G \rho_0}
\end{equation} 
is the dimensionless parameter that characterizes the rotation strength and 
\begin{equation}\label{Psi}
\Psi\equiv\psi^{(int)}({\bf 0})
\end{equation} 
is the depth of the potential well at the center. The dimensionless density profile is given by 
\begin{equation}\label{rho} 
\hat{\rho}_K(\psi)= \frac{\rho_K(\psi)}{\hat{A}}=\mathrm{e}^{\psi}\gamma\left(\frac{5}{2},\psi\right)~, 
\end{equation}
where 
\begin{equation}\label{hatA}
\hat{A} = \frac{8 \pi2^{1/2} A }{3 a^{3/2}}\mathrm{e}^{-a H_0}
\end{equation}
and $\gamma$ denotes the incomplete gamma function. Therefore, the central density is given by $\rho_0=\hat{A}\hat{\rho}_K(\Psi)$. 
Outside the cluster, for negative values of $\psi$, we should refer to the Laplace equation
\begin{equation}\label{Laplace}
\hat{\nabla}^2 \psi^{(ext)}=18 \chi~.
\end{equation}

The relevant boundary conditions are given by the requirement of regularity of the solution at the origin and by the condition that $\psi^{(ext)}+ a \Phi_{cen}\rightarrow aH_0$ at large radii. The Poisson (internal) and Laplace (external) domains are thus separated by the boundary surface, which is unknown {\em a priori}. Therefore, we have to solve an elliptic partial differential equation in a free boundary problem. In particular, here we illustrate the properties of the solutions of the Poisson-Laplace equation obtained by using a perturbation method, which also requires an expansion of the solution in Legendre series. To obtain a uniformly valid solution over the entire space, an asymptotic matching is performed between the internal and the external solution, using the Van Dyke principle (see Van Dyke \cite{Dyk75}). This method of solution is basically the same as proposed by Smith (\cite{Smi75}) for the construction of rotating configurations with polytropic index $n=3/2$. We have calculated the complete solution up to second-order in the rotation strength parameter $\chi$. The final solution is expressed in spherical coordinates ${\bf \hat{r}}=(\hat{r}, \theta, \phi)$ and the resulting configurations are characterized by axisymmetry (i.e., the density distribution and the potential do not depend on the azimuthal angle $\phi$). For the details of the method for the construction of the solution the reader is referred to Paper I and to Varri (\cite{thesis}), in which the complete calculation is provided.

\begin{figure}[t!]
\centering
  \includegraphics[height=.3\textheight]{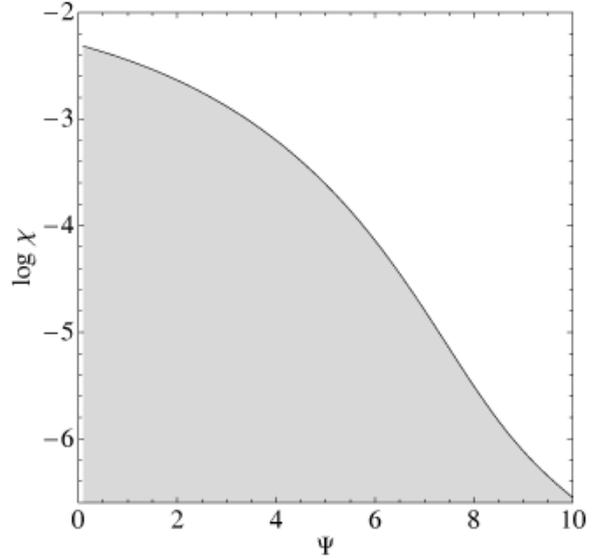}
  \caption{\normalsize Parameter space for second-order rigidly rotating models. 
  The solid line represents the critical values of the rotation strength 
  parameter $\chi_{cr}$ and the grey region identifies the 
  values $(\Psi,\chi)$ for which the resulting models are bounded by a 
  closed constant-$\psi$ surface (subcritical models).}
  \label{parspace}
\end{figure}
	

\subsection{The parameter space}
\label{Unif2}

The resulting models are characterized by two dimensional scales (e.g., the total mass and the core radius) and two dimensionless parameters. As in the spherical King models, the first parameter measures the \emph{concentration} of the configuration; we thus consider the quantity\footnote{In the literature, the parameter $\Psi$ is often denoted by $W_0$; here we prefer to keep the same notation used in Paper I, Paper II, and in other articles.} $\Psi$ (see Eq.~(\ref{Psi})) or equivalently $c\equiv \log(r_{tr}/r_0)$, where $r_{tr}$ is the truncation radius of the spherical King model associated with a given value of $\Psi$. The second dimensionless parameter $\chi$ (see Eq.~(\ref{chi})) characterizes the \emph{rotation strength} measured in terms of the frequency associated with the central density of the cluster.

For every value of the dimensionless central concentration $\Psi$ there exists a maximum value of the rotation strength parameter, corresponding to a critical model, for which the boundary is given by the critical constant-$\psi$ surface. The boundary surface of a representative critical model (with $\Psi=2$) is depicted in Fig.~\ref{crit}; the surface is such that all the points on the equatorial plane are saddle points, where the centrifugal force balances the self-gravity. We refer to their distance from the origin as $\hat{r}_B$, the \emph{break-off radius}. In the constant-$\psi$ family of surfaces associated with a given value of $\Psi$, the critical surface thus separates the open from the closed surfaces. Consistent with the assumption of stationarity, only configurations bounded by closed surfaces are considered here. This unique geometrical characterization suggests that the effect of the rotation may be expressed also in terms of an {\em extension} parameter
\begin{equation}
\delta=\frac{\hat{r}_{tr}}{\hat{r}_{B}}~,
\end{equation}
which provides an indirect measure of the deviations from sphericity of a configuration, by considering the ratio between the truncation radius of the corresponding spherical King model and the break-off radius of the associated critical surface. Therefore, a given model may be labelled by the pair $(\Psi,\chi)$ or equivalently by the pair $(\Psi, \delta)$. For given $\Psi$, there is thus a maximum value of the allowed rotation, which we may express as $\chi_{cr}$ or $\delta_{cr}$. A model with $\delta < \delta_{cr}$ may be called subcritical.

For each value of $\Psi$, the critical value of the rotation parameter can be found by numerically solving the system\footnote{For brevity, here we omit the explicit structure of the system. The reader is referred to Eqs.~(8)-(14) in Paper II, with respect to which several differences occur, because in the axisymmetric rotation problem the coefficients of the asymptotic series are best expanded in (normalized) Legendre polynomials (rather than spherical harmonics). In particular, with respect to Paper II: (i) the coefficients with $m\ne 0$ must be dropped; (ii) the coefficients with $l\ne0$ must be multiplied by the factor $(\pi/4)^{1/2}$, because of the different normalization used in the two systems of orthonormal functions; (iii) the expressions are evaluated at $\hat{r}_B$ instead of $\hat{r}_T$. As in Paper I, for the definition of the Legendre polynomials we refer to Eqs.~(22.3.8) and (22.2.10) of Abramowitz \& Stegun (\cite{AbrSte65}).}
\begin{equation}\label{sys}
\left\{
\begin{array}{ll}\label{sistcrit}
\partial_{\hat{r}}\, \psi(\hat{r}=\hat{r}_B,\theta=\pi/2;\chi_{cr})=0\vspace{.1cm}\\
\psi(\hat{r}=\hat{r}_B,\theta=\pi/2;\chi_{cr})=0~,
\end{array}
\right.
\end{equation}  
where the unknowns are $\hat{r}_B$ and $\chi_{cr}$. In terms of the extension parameter, for a given $\Psi$, the critical condition occurs when $\delta=\delta_{cr}=\hat{r}_{tr}/\hat{r}_B\approx 2/3$. This value is obtained by inserting in Eq.~(\ref{sys}) the zeroth-order expression for the cluster potential, as discussed in detail in Sect.~2 of Paper II for the tidal problem (see p.~251-255 in Jeans \cite{Jea29} for an equivalent discussion referred to the purely rotating Roche model, that is a rotating configuration in which a small region with infinite density is surrounded by an ``atmosphere'' of negligible mass). 

\begin{figure}[t!]
\centering
  \includegraphics[height=.3\textheight]{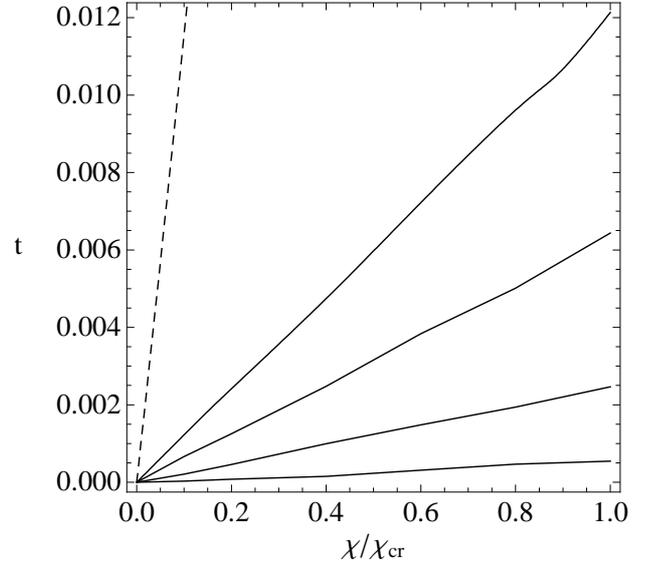}
  \caption{ \normalsize Values of the ratio between ordered kinetic energy and gravitational 
energy $t=K_{ord}/|W|$ for selected rigidly rotating models characterized by $\Psi=1,3,5,7$ (solid lines, from top to bottom) and $\chi$ in the range $[0,\chi_{cr}]$. For comparison, the dashed line indicates the values of $t$ for the sequence of Maclaurin oblate spheroids in the limit of small eccentricity.}
  \label{tunif}
\end{figure}

The parameter space for the second-order models is presented in Fig.~\ref{parspace} (which corresponds to Fig.~1 of Paper II describing the tidal models). Two rotation regimes exist, namely the regime of low-deformation ($\delta \ll \delta_{cr}$, bottom left corner), where internal rotation does not affect significantly the morphology of the configuration, which remains very close to spherical symmetry, and that of high-deformation ($\delta \approx \delta_{cr}$, close to the solid line), where the model is highly affected by the nearly critical rotation velocity, especially in the outer parts. Note that the actual regime depends on the combined effect of rotation strength and of concentration. In other words, the models described here belong to the class of rotating configurations characterized by equatorial break-off (``region of equatorial break-off'', see Fig.~44, p.~267 in Jeans \cite{Jea29}), for which the limiting case is given by the purely rotating Roche model.     

A global kinematical characterization, complementary to the information provided by the rotation strength parameter, is offered by the parameter $t=K_{ord}/|W|$, defined as the ratio between ordered kinetic energy and gravitational energy. Figure~\ref{tunif} illustrates the relation between the two parameters for models with selected values of $\Psi$ and increasing values of $\chi$, up to the critical configuration characterized by $\chi_{cr}$. The parameter $t$ increases linearly for increasing values of $\chi$ (with a slope dependent on the concentration parameter $\Psi$). A similar linear relation is observed for small values of eccentricity ($e << 1$) in the sequence of Maclaurin oblate spheroids $t(e) \sim \chi(e) \sim 2 e^2 /15$ (for the definitions of the two parameters in the context of Maclaurin spheroids, see Eqs.~(10.20) and (10.24) in Bertin \cite{Ber00}); in Fig. \ref{tunif} the relevant linear relation is normalized with respect to the maximum value of the rotation strength parameter attained in the sequence of Maclaurin spheroids $\chi_{max}=0.11233$ (see Eq.~(10), p.~80 in Chandrasekhar \cite{Cha69}).

\subsection{Intrinsic properties}
\label{Unif3}

\begin{figure}[t!]
  \includegraphics[height=.29\textheight]{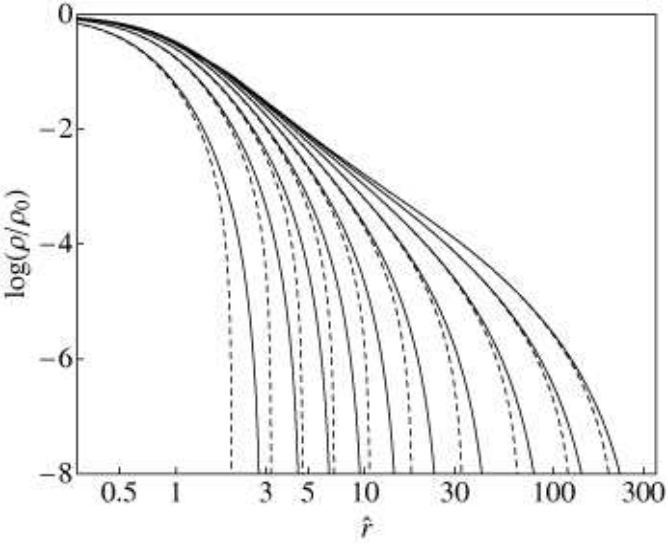}
  \caption{ \normalsize Intrinsic density profiles (normalized to the 
  central value) evaluated on the equatorial plane (solid lines) and along the 
  $\hat{z}$-axis (dashed lines) for critical second-order rigidly
  rotating models with $\Psi=1,2,...,10$ (from left to right).}
  \label{unif_dens}
\end{figure}

The geometry of the models, reflecting the properties of the centrifugal potential, is characterized by symmetry around the $\hat{z}$-axis and reflection symmetry with respect to the equatorial plane $(\hat{x},\hat{y})$. As expected, compared to the corresponding spherical King models, the rotating models stretch out on the equatorial plane and are slightly flattened along the direction of the rotation axis (see Fig.~\ref{unif_dens} for the density profiles of selected critical second order models evaluated on the equatorial plane and along the $\hat{z}$-axis). In general, configurations in the low-deformation regime ($\delta \ll \delta_{cr}$), regardless of the value of concentration $\Psi$, are almost indistinguishable from the corresponding spherical King models. 

\begin{figure}[t!]
\centering
  \includegraphics[height=.29\textheight]{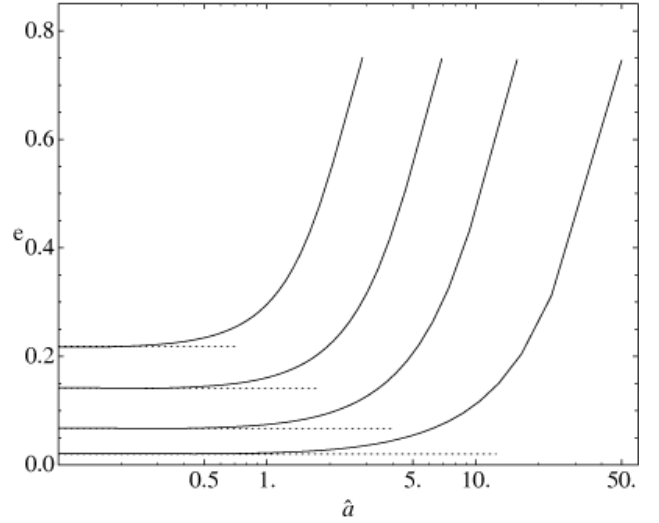}
  \caption{ \normalsize Eccentricity profiles 
  of the isodensity surfaces for selected critical second-order rigidly rotating models, 
  with $\Psi=1,3,5,7$ (from top to bottom); dotted horizontal lines show the central 
  eccentricities values estimated analytically (see Eq.~(\ref{e0})).\vspace{0.2cm}}
  \label{ecce}
\end{figure}

Models in the intermediate and high-deformation regime ($\delta \approx \delta_{cr}$) show modest deviations from spherical symmetry in the central regions, while they are significantly flattened in the outer parts. The intrinsic eccentricity profile, defined as $e=[1-(\hat{b}/\hat{a})^2]^{1/2}$, where $\hat{a}$ and $\hat{b}$ are semi-major and semi-minor axes of the isodensity surfaces, is a monotonically increasing function of the semi-major axis (see Fig.~\ref{ecce} for the computed eccentricity profiles of selected critical second-order models). We recall that the geometry of the boundary surface of a critical model depends only slightly on the value of the concentration parameter. In particular, in the critical case, the break-off radius $\hat{r}_B$ represents the distance from the center of the outermost points of the boundary surface on the equatorial plane and the truncation radius $\hat{r}_{tr}$ is approximately the distance of the last point on the polar axis (i.e., the $\hat{z}$-axis). Therefore, since $\delta_{cr}= \hat{r}_{tr}/\hat{r}_B\approx 2/3$, the value of the termination points of the eccentricity profiles of critical models is approximately the same ($e \approx 0.75$; see the termination points of the solid lines in Fig.~\ref{ecce}). 

In addition, by using the multipolar structure of the solution of the Poisson-Laplace equation obtained with the perturbation method, the asymptotic behavior of the eccentricity profiles in the central regions can also be evaluated analytically. Since the distribution function depends only on the Jacobi integral (i.e., the isolating energy integral in the corotating frame of reference), the density and the velocity dispersion profiles are functions of only the escape energy (see Eqs.~(\ref{rho}) and (\ref{sigma}), respectively). Therefore, there is a one-to-one correspondence between equipotential, isodensity, and isobaric surfaces and the eccentricity profiles can be calculated with reference to just one of these families of surfaces. In fact, if expanded to second order in the dimensionless radius, the escape energy in the internal region reduces to:
\begin{eqnarray}
\psi^{(int)}(\hat{r},\theta)&=& \Psi +\frac{1}{2}\left[- 3 +  6 \chi+ 
2A_2 U_2(\theta) \right.\chi+\nonumber\\
&&\left. (B_2+1) U_2(\theta) \chi^2\right]\hat{r}^2+ \mathcal{O}(\hat{r}^4)~,
\end{eqnarray}
where $U_2(\theta)$ denotes the normalized Legendre polynomial with $l=2$ and $A_2, B_2$ are appropriate (negative) coefficients, depending on $\Psi$,  which are determined by asymptotically matching the internal and external solution, in order to have continuity on the entire domain
(see Eqs.~(62) and (68) in Paper I, to be interpreted as indicated in Appendix B of Paper I). By setting $\psi^{(int)}(\hat{a},\pi/2)=\psi^{(int)}(\hat{b},0)$, we thus find that, in the innermost region, the eccentricity tends to the following nonvanishing central value
\begin{equation}\label{e0}
e_0=\frac{[6 A_2\,\chi + 3(B_2+1)\,\chi^2]^{1/2}}{[6 \sqrt{2/5}(2\chi-1)+4 A_2\,\chi+2(B_2+1)\,\chi^2]^{1/2}}~.
\end{equation}
Therefore, the central value of the eccentricity is finite, of order $\mathcal{O}(\chi^{1/2})$, and strictly vanishes only in the limit of vanishing 
rotation strength. This result is nontrivial because the centrifugal potential (which induces the deviations from sphericity) is a homogeneous function of the spatial coordinates. Therefore, we might naively expect that, in their central regions, the models reduce to a perfectly spherical shape (i.e., $e_0 = 0$), even for finite values of the rotation strength. This property has been noted also in the family of triaxial tidal models, in which the tidal potential plays the role of the centrifugal potential (see Sect.~3.1 of Paper II).      

Deviations from spherical symmetry can also be described, in a global way, by the quadrupole moment tensor, defined as
\begin{eqnarray}\label{quad}
Q_{ij}&=&\int_V\,(3x_ix_j-r^2\delta_{ij})\rho({\bf r}) d^3r= \nonumber \\
&&\hat{A}r_0^5\int_V\,(3\hat{x}_i\hat{x}_j-\hat{r}^2\delta_{ij})
\hat{\rho}({\bf \hat{r}}) d^3\hat{r}=\hat{A}r_0^5\hat{Q}_{ij}~,
\end{eqnarray}
where the integration is performed in the volume $V$ of the entire configuration. It can be easily shown that in our coordinate system the tensor is diagonal and that $Q_{xx}=Q_{yy}$. The components of the tensor can be calculated explicitly
\begin{equation}\label{quad}
\hat{Q}_{xx}^{(2)}=\hat{Q}_{yy}^{(2)}=-\frac{\hat{Q}_{zz}^{(2)}}{2}=2\pi 
\frac{\sqrt 10}{9}\hat{\rho}_K(\Psi)\left(a_2\chi+b_2 \frac{\chi^2}{2}\right)~,
\end{equation}
\begin{figure}[t!]
  \includegraphics[height=.29\textheight]{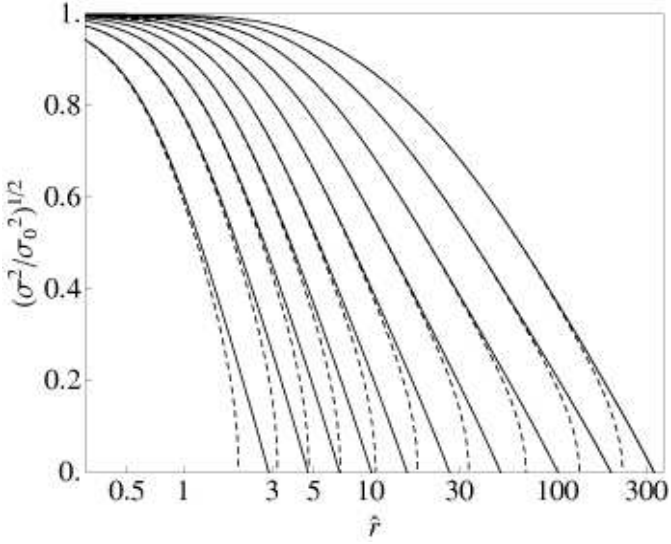}
  \caption{\normalsize Intrinsic velocity dispersion profiles (normalized to the 
  central value) evaluated on the equatorial plane (solid lines) and 
  along the $\hat{z}$-axis (dashed lines) for selected critical second-order rigidly rotating 
  models with $\Psi=1,2,...,10$ (from left to right).}
  \label{unif_vel}
\end{figure}

The quantities $a_2$ and $b_2$ are appropriate (positive) coefficients, depending on $\Psi$, resulting from the asymptotic matching of the internal and external solution of the Poisson-Laplace equation. The sign of the components are consistent with the above mentioned compression and stretching of the density distribution.  The expression in Eq.~(\ref{quad}) is calculated from the second-order external solution; the first-order expression is recovered by dropping the quadratic term in the parameter $\chi$. At variance with the tidal case, the ratio $\hat{Q}_{zz}/\hat{Q}_{xx}$ is independent of the rotation parameter $\chi$. This analytical result has been compared to the ratio of the components of the quadrupole tensor determined by direct numerical integration (performed by means of the algorithm VEGAS, see Press et al. \cite{Pre92} ) and good agreement has been found. \footnote{Strictly speaking, the analytical expressions for the components of the quadrupole moment tensor refer only to the inner region, because the contribution from the boundary layer, where $\psi $ is of order $\mathcal{O}(\chi)$, is neglected.} For the tidal case, the detailed calculation can be found in Sect.~3.3 and Appendix B of Paper II; the calculation is easily adapted to the rotating case.  

By construction, the models are isotropic in velocity space, with the dimensionless scalar velocity dispersion given by 
\begin{equation}\label{sigma}
\hat{\sigma}_K^2(\psi)=\frac{2}{5}\frac{\gamma(7/2,\psi)}{\gamma(5/2,\psi)}~.
\end{equation}
As noted for the intrinsic density profiles, a compression along the vertical axis and a stretching along the equatorial plane occur also for the velocity 
dispersion profiles (the profiles of selected critical second-order models are shown in Fig.~\ref{unif_vel}).

\begin{figure}[t!]
\centering
  \includegraphics[height=.29\textheight]{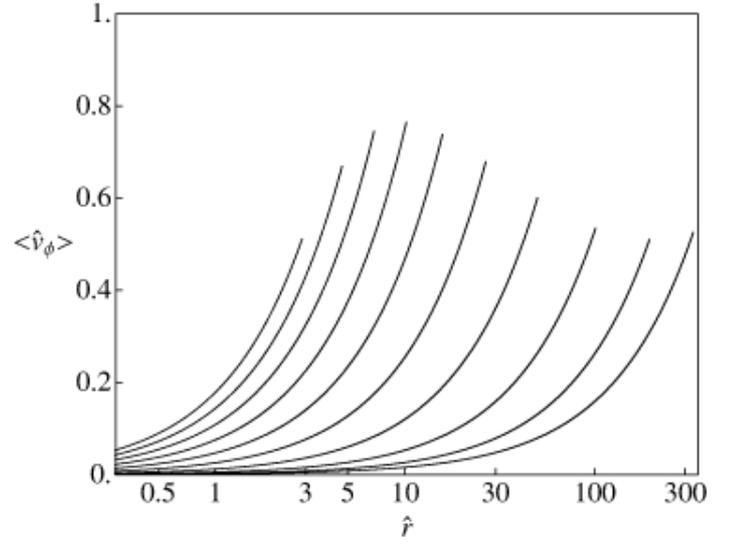}
  \caption{ \normalsize Mean rotation velocity profiles on the equatorial plane for the selected 
  critical second-order rigidly rotating models with $\Psi=1,2,...,10$ (from left to right). The mean velocity increases linearly with radius because the rotation is rigid (see Eq.~(\ref{velunif}); note the logarithmic scale of the horizontal axis). The values of the termination points of the curves depend on the value of $\chi_{cr}$ (which decreases as $\Psi$ increases, see Fig.~\ref{parspace}) and on the extension of the models on the equatorial plane (which increases as $\Psi$ increases, see Fig.~\ref{unif_dens}). }
  \label{vel}
\end{figure}

The mean rotation velocity (which is subtracted away when the corotating frame of reference is considered) 
characterizing the models is defined as $\langle{\bf v}\rangle= \omega \,\hat{{\bf e}}_{z} \times r \,\hat{{\bf e}}_{r}= \langle v_{\phi}\rangle\hat{{\bf e}}_{\phi}$; in the adopted dimensionless units, the azimuthal component can be written as  
\begin{equation}\label{velunif}
\langle\hat{v}_{\phi}\rangle=\langle v_{\phi}\rangle a^{1/2}=3 \chi^{1/2} \hat{r} \sin{\theta}~.
\end{equation}
As expected, the mean velocity is constant on cylinders (in our coordinate system, the cylindrical radius is defined by $\hat{R}=\hat{r}\sin{\theta}$). The relevant dimensionless angular velocity is linked to the rotation strength parameter by the following relation $\hat{\omega}=3 \chi^{1/2}$ (the numerical factor $3$ is due to the adopted scale length, see Eq.~(\ref{r0})). The rotation profiles of selected second-order critical models are represented in Fig.~\ref{vel}. 
For all the models, as we approach the boundary of the configuration, the ratio $\langle\hat{v}_{\phi}\rangle/\hat{\sigma}_K$ quickly diverges since at the boundary the rotation velocity tends to a finite value while the velocity dispersion vanishes; this behavior is observed in every direction (except for the $\hat{z}$-axis, on which the rotation is absent by definition).

\begin{figure*}[t!]
\centering
  \includegraphics[height=.45\textheight]{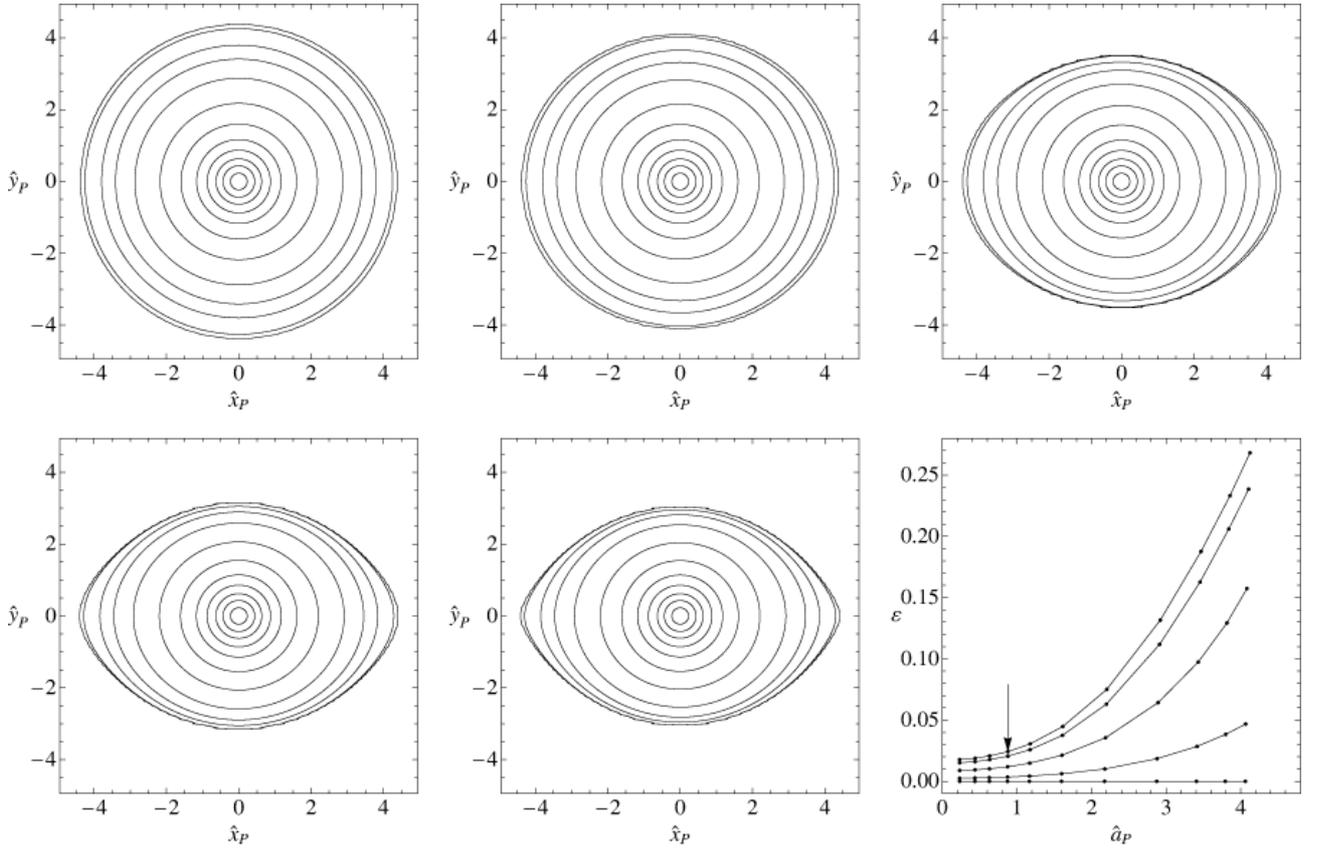}
  \caption{ \normalsize Projections along directions identified by $\phi=0$ and 
  $\theta=i(\pi/8)$ with $i=0,..,4$ (from left to right, top to bottom) 
  of a critical second-order rigidly rotating model with $\Psi=2$; 
  the first and the fifth panel represent the 
  projections along the $\hat{z}$-axis (``face on'') and the $\hat{x}$-axis 
  (``edge-on'') of the intrinsic coordinate system, respectively. Solid lines mark 
  the isophotes, corresponding to selected values of $\Sigma/\Sigma_0$ in 
  the range $[0.9, 10^{-7}]$. The last panel shows the ellipticity profiles, 
  as functions of the semi-major axis of the projected image $\hat{a}_P$, 
  referred to the lines of sight considered in the previous panels (from 
  bottom to top). Dots represent the locations of the isophotes and the 
  arrow marks the position of the half-light isophote (practically the 
  same for every projection considered in the figure).}
  \label{proj_unif}
\end{figure*}

\subsection{Projected properties}
\label{Unif4}

For a comparison of the models with the observations (under the assumption of a constant mass-to-light ratio), we have then
computed surface (projected) density profiles and isophotes. The projection has been performed along selected directions, identified by the viewing angle $(\theta,\phi)$ corresponding to the $\hat{z}_P$ axis of a new coordinate system related to the intrinsic system by the transformation ${\bf
\hat{x}}_P=R{\bf \hat{x}}$; the rotation matrix $R=R_1(\theta)R_3(\phi)$ is expressed in terms of the viewing angles, by taking the $\hat{x}_P$ axis as the line of nodes (i.e., the same projection rule we adopted for the triaxial tidal models, see Sect.~3.4 in Paper II). Since the rigidly rotating models are characterized by axisymmetry with respect to the $\hat{z}$-axis and reflection symmetry with respect to the equatorial plane, it is sufficient to choose 
the viewing angles from the $(\hat{x},\hat{z})$-plane of the intrinsic coordinate system. In particular, we used the line of sights defined by $\theta_i=i(\pi/8)$ and $\phi=0$ with $i=0,..,4$, and we calculated (by numerical integration, using the Romberg's rule) the dimensionless projected density
\begin{equation}\label{Sigma}
\hat{\Sigma}(\hat{x}_P,\hat{y}_P)=\int_{-\hat{z}_{sp}}^{\hat{z}_{sp}}
\hat{\rho}({\bf \hat{r}}_P)d\hat{z}_P~,
\end{equation}
where $\hat{z}_{sp}=(\hat{x}_\mathrm{e}^2-\hat{x}_P^2-\hat{y}_P^2)^{1/2}$ with $\hat{x}_e$ the edge of the model along the $\hat{x}$ axis of the intrinsic coordinate system. The projection plane $(\hat{x}_P,\hat{y}_P)$ has been sampled on an equally-spaced square cartesian grid centered at the origin.
 
The first five panels of Fig.~\ref{proj_unif} show the projected images of a critical second-order model with $\Psi=2$; the first and the fifth panel correspond, respectively, to the least and to the most favorable line of sight for the detection of the intrinsic flattening of the model (the $\hat{z}$ and $\hat{x}$-axis of the intrinsic coordinate system, that is ``face-on'' and ``edge-on'' view). 

The morphology of the isophotes of a given projected image can be described in terms of the {\it ellipticity} profile, defined as $\varepsilon=1-\hat{b}_P/\hat{a}_P$ where $\hat{a}_P$ and $\hat{b}_P$ are the principal semi-axes, as a function of the semi-major axis $\hat{a}_P$. As already noted for the (intrinsic) eccentricity profile, the deviation from circularity increases with the distance from the origin. In the inner region, the central value of the ellipticity is consistent with the central eccentricity $e_0$ calculated in the previous subsection. The last panel of Fig.~\ref{proj_unif} illustrates the ellipticity profiles corresponding to the projections displayed in the previous panels. In addition, the isophotes of models in the high deformation regime ($\delta \approx \delta_{cr}$), if projected along appropriate line of sights, show clear departures from a pure ellipse, that can be characterized as a ``disky'' overall trend (e.g., see Jedrzejewski \cite{Jed87}), particularly evident in the outer parts (see fourth and fifth panels in Fig.~\ref{proj_unif}).    
 
Because the family of rigidly rotating models is characterized by simple kinematical properties (pressure isotropy and solid-body rotation), that have been already presented in detail with reference to three-dimensional configurations, for brevity, the derivation of the projected kinematical properties is omitted here; a full description can be found in Varri (\cite{thesis}). 

\begin{table*}
    \caption[]{\normalsize Summary of the properties of the families of models studied in the present paper.}
       \label{scheme}
\begin{center}
       \begin{tabular}{ccccccc}
          \hline
          \hline
          \noalign{\smallskip}
          Family of & Distribution   & Dimensionless  & Internal & Anisotropy  & Isophote & Paper\\
          models & function & parameters & rotation & profile & shape& sections \\
          \noalign{\smallskip}
          \hline
          \noalign{\smallskip}
		$\;\;f_K^r(H)$		& 	$A\mathrm{e}^{-aH_0}[\mathrm{e}^{-a(H-H_0)}-1]$	&	$\Psi$, $\chi$	& solid-body & $(0,0,0)$ & disky& 2 \\ \noalign{\smallskip}
		$\;\;f_{WT}^d(I)$		&	$A \mathrm{e}^{-aE_0}[\mathrm{e}^{-a(I-E_0)} - 1 + a(I-E_0)]$	&	$\Psi$, $\chi$, $\bar{b}$, $c$	& differential& $(0,>0,-2)$ & boxy & 3,\,4,\,5,\,6 \\ \noalign{\smallskip}
		 $[~f_{PT}^d(I)$	&	$A \mathrm{e}^{-aE_0}\mathrm{e}^{-a(I-E_0)}$	&	$\Psi$, $\chi$, $\bar{b}$, $c$ & differential & $(0,>0,0)$ & boxy& 3,\,6,\,App.~B\,$]$ \\	
          \noalign{\smallskip}
          \hline
       \end{tabular}
       \tablefoot{The relevant integrals are defined as $H = E- \omega J_z$ and $I=E -\omega J_z/(1 +b J_z^{2\,c})$, with the corresponding cut-off constants given by $H_0$ and $E_0$. The dimensionless parameters are defined as follows: $\Psi=\psi({\bf 0})$ respresents a measure of the concentration, $\chi=\omega^2/(4 \pi G \rho_0)$ a measure of the (central) rotation strength, and, for the family of differentially rotating models, $\bar{b}=b r_0^{2c}a^{-c}$ and $ c > 1/2$ determine the shape of the rotation profile. The pressure anisotropy profiles are characterized in terms of the values of the anisotropy parameter $\alpha=1- \sigma^2_{\phi\phi}/\sigma^2_{rr}$ in the central, intermediate, and outer regions of a model; values of $\alpha$ greater than, lower than, and equal to zero indicate radially-biased, tangentially-biased anisotropy, and isotropy in velocity space, respectively.  For each family of models, the first and third values of $\alpha$ are calculated analytically as the limiting values for small and large radii. For physical reasons discussed in Sect.~3.1, the focus of the paper is on the first two families.}
\end{center}
\end{table*}
 
\section{Differentially rotating models}
\label{Diff}
\subsection{Choice of the distribution function}
\label{Diff1}

As indicated in the Introduction, theoretical and observational motivations have brought us to look for more realistic configurations, characterized by differential rotation. Thus we focus our attention on axisymmetric systems, within the class of distribution functions that depend only on the energy $E$ and the z-component of the angular momentum $J_z$, and we consider the integral
\begin{equation} \label{I}
I(E,J_z)= E -\frac{\omega J_z}{1 +b J_z^{2\,c}}~,
\end{equation}
where $\omega$, $b$,  and $c > 1/2$ are positive constants. The quantity $I(E,J_z)$ reduces to the Jacobi integral for small values of the z-component of the angular momentum and tends to the single-star energy in the limit of high values of $J_z$. Therefore, if we refer to a distribution function of the form $f = f(I)$, we may argue that $\omega$ is related to the angular velocity in the central region of the system, characterized by approximately solid-body rotation, whereas the positive constants $b,c$ will determine the shape of the radial profile of the rotation profile. In view of the arguments that have led to the truncation prescription that characterizes King model, we decided to introduce a truncation in phase-space based exclusively on the single-star energy with respect to a cut-off constant $E_0$.   

For simplicity, we consider two families of distribution functions. The first family is defined as 
\begin{equation}\label{fW}
f_{WT}^d(I)= A \mathrm{e}^{-aE_0}\left[\mathrm{e}^{-a(I-E_0)} - 1 + a(I-E_0)\right]
\end{equation}
if $E \le E_0$ and $f_{WT}^d(I) = 0$ otherwise, so that both $f_{WT}^d(I)$ and its derivative with respect to $E$ are continuous. We refer to this truncation prescription as Wilson truncation (hence the subscript $WT$) because, in the limit of vanishing internal rotation ($\omega \rightarrow 0$), this family reduces to the spherical limit of the distribution function proposed by Wilson (\cite{Wil75}). The superscript $d$ in Eq.~(\ref{fW}) indicates the presence of differential rotation.  

The second family is defined by the distribution function 
\begin{equation}\label{fPT}
f_{PT}^d(I)= A \mathrm{e}^{-aE_0}\mathrm{e}^{-a(I-E_0)}
\end{equation}
if $E \le E_0$ and $f_{PT}^d(I) = 0$ otherwise; therefore the function, characterized by plain truncation (hence the subscript $PT$), is discontinuous with respect to $E$. In the limit of vanishing internal rotation, it reduces to the spherical limit of the function proposed by Prendergast \& Tomer (\cite{PreTom70}), which leads to the truncated isothermal sphere (see also Woolley \& Dickens \cite{WooDic62}). A summary of the main properties and definitions of the relevant nonrotating limit of the two families of models is presented in Appendix \ref{AppA}. 

In both cases the distribution functions are positive definite $f_{WT}^d(I),\,f_{PT}^d(I) \ge 0$, by construction. Curiously, a naive extension of King models $f = f_K^d(I)$ with a similar truncation in energy alone would define a distribution function that is not positive definite in the whole domain of definition. 

Sharp gradients or discontinuities in phase space (such as the ones associated with the truncation prescription of $f_{PT}^d(I)$) are expected to be associated with evolutionary processes dictated either by collective modes or by any small amount of collisionality. Therefore, the first truncation prescription, corresponding to a smoother distribution in phase space, is to be preferred from a physical point of view as the basis for a realistic equilibrium configuration (in principle, we might have referred to even smoother functions; see Davoust \cite{Dav77}). In addition, a full analysis of the configurations defined by $f_{WT}^d(I)$ shows that this family of models exhibits a number of interesting intrinsic and projected properties, more appropriate for application to globular clusters, with respect to the models defined by $f_{PT}^d(I)$.  

Therefore, the following Sects. \ref{DiffIntr} and \ref{DiffProj} are devoted to the full characterization of the family of models defined by $f_{WT}^d(I)$ (for a summary of the properties of the families of models studied in the present paper, see Table 1). The intrinsic properties of the family of models defined by $f_{PT}^d(I)$ are summarized in Appendix B. In this investigation, we decided to briefly mention and to keep also the second family not only because it extends a well-known family of models, but also because it allows us to check directly an important aspect of model construction that had been noted by Hunter (\cite{Hun77}). This is that the truncation prescription affects the density distribution in the outer parts of the models significantly. This general point is even more important if we are interesting in modeling the outermost regions of globular clusters (and the issues that are often discussed under the name of studies of ``extra-tidal light", e.g., see Jordi \& Grebel \cite{Jor10}). 

\subsection{The construction of the models}
\label{Diff2}
The construction of the models requires the integration of the relevant Poisson equation, supplemented by a set of boundary conditions equivalent to the one described in Sect.~2 for rigidly rotating models.  In this case, we obtain the solution by means of an iterative approach, based on the method proposed by Prendergast \& Tomer (\cite{PreTom70}), in which an improved solution $ \psi^{(n)}$ of the dimensionless Poisson equation is obtained by evaluating the source term on the right-hand side with the solution from the immediately previous step (for an application of the same method to the construction of configurations shaped by an external tidal field, see Sect.~5.2 in Paper I)
\begin{equation}
\hat{\nabla}^2 \psi^{(n)}=-\frac{9}{\hat{\rho}_0}\hat{\rho}\left(\hat{r},\theta,\psi^{(n-1)}\right)~;
\end{equation}
here the dimensionless escape energy is 
\begin{equation}\label{psidiff}
\psi({\bf r})=a[E_0-\Phi_C({\bf r})]~,
\end{equation}
the dimensionless radius is defined as $\hat{r}=r/r_0$, with the same scale length introduced in Eq.~(\ref{r0}), and $\hat{\rho}_0$ indicates the dimensionless central density. The relevant density profile $\rho=\hat{A}\hat{\rho}$, with $\hat{A}$ defined as in Eq.~(\ref{hatA}), results from the integration in velocity space of the distributions function defined by $f_{WT}^d(I)$. It is clear that the general strategy for the construction of the self-consistent solution is applicable also to the density derived from $f_{PT}^d(I)$.
\begin{figure}[t!]
\centering
\includegraphics[height=.3\textheight]{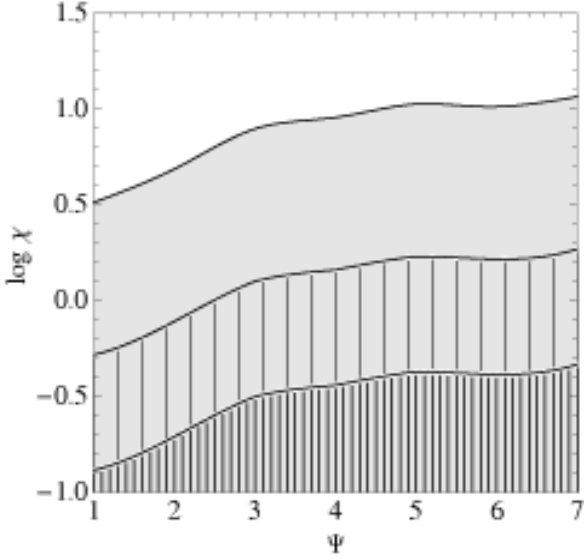}
\caption{ \normalsize Two-dimensional parameter space, given by 
central rotation strength $\chi$ vs. concentration $\Psi$, of differentially rotating models defined by $f_{WT}^d(I)$; the remaining 
parameters are fixed at $\bar{b}=c=1$. The upper solid line marks 
the maximum admitted values of $\chi$, for given 
values of concentration $\Psi$, that is the underlying area indicates 
pairs $(\Psi,\chi)$ for which models can be constructed. The intermediate and the lower solid 
lines mark the values of $\hat{\omega}/\hat{\omega}_{max}=0.4,0.2$, respectively. 
The gray, wide-striped, and thin-striped areas represent the extreme, rapid, 
and moderate rotation regimes, respectively.}
\label{parspacediff}
\end{figure}
The iteration is seeded by the corresponding spherical models, that is, the Wilson and the Prendergast-Tomer spheres respectively, and is stopped when numerical convergence is reached (see Appendix \ref{AppC} for details). At each iteration step, the scheme requires the expansion in Legendre series of the density and the potential
\begin{equation}\label{psin}
\psi^{(n)}({\bf \hat{r}
})=\sum_{l=0}^{\infty}\psi_{l}^{(n)}(\hat{r})U_{l}(\cos \theta)~,
\end{equation}
\begin{equation}\label{rhon}
\hat{\rho}^{(n)}({\bf \hat{r}
})=\sum_{l=0}^{\infty}\hat{\rho}_{l}^{(n)}(\hat{r})U_{l}(\cos \theta)~.
\end{equation}

The associated Cauchy problems for the radial functions $\psi_{l}^{(n)}(\hat{r})$ are therefore
\begin{equation}\label{Cauchy}
\left[\frac{d^2 }{ d \hat{r}^2}+\frac{2}{\hat{r}}{d \over d
\hat{r}} -{l(l+1)\over \hat{r}^2}
\right]\psi_{l}^{(n)}=-\frac{9}{\hat{\rho}_0}
\hat{\rho}_{l}^{(n-1)}~,
\end{equation}
supplemented by the following boundary conditions
\begin{equation}\label{BC1}
\psi_{0}^{(n)}(0)=\Psi\sqrt{2}~,
\end{equation}
\begin{equation}\label{BC2}
\psi_{l}^{(n)}(0)=0~, \mbox{for $l\ne0$}
\end{equation}
\begin{equation}\label{BC3}
{\psi_{0}^{(n)}}'(0)={\psi_{l}^{(n)}}'(0)=0~,
\end{equation}
where $\Psi$ is the depth of the dimensionless potential well at the center. By using the method of variation of arbitrary constants, the radial functions can be expressed in integral form as follows
\begin{eqnarray}\label{eqint0}
\psi_{0}^{(n)}(\hat{r})&=&\Psi\sqrt{2} -\frac{9}{\hat{\rho}_0}\left[
\int_0^{\hat{r}}\hat{r}'\hat{\rho}_{0}^{(n-1)}
(\hat{r}')d\hat{r}'\right.\nonumber \\
&&\left. -\frac{1}{\hat{r}}\int_0^{\hat{r}}
\hat{r}'^2\hat{\rho}_{0}^{(n-1)}(\hat{r}')d\hat{r}'\right]~,
\end{eqnarray}
\begin{eqnarray}\label{eqintn}
\psi_{l}^{(n)}(\hat{r})&=&\frac{9}{(2l+1)\hat{\rho}_0}\left[\hat{r}^{l}
\int_{\hat{r}}^{\infty}
\hat{r}'^{1-l}\hat{\rho}_{l}^{(n-1)}(\hat{r}')d\hat{r}'\right. \nonumber \\
&& \left. +\frac{1}{\hat{r}^{l+1}}\int_0^{\hat{r}}
\hat{r}'^{l+2} \hat{\rho}_{l}^{(n-1)}(\hat{r}')d\hat{r}'\right]~.
\end{eqnarray}
The factor $\sqrt{2}$ appearing in Eqs.~(\ref{BC1}) and (\ref{eqint0}) is due to the normalization assumed for the Legendre polynomials. 

\subsection{The parameter space}
\label{Diff3}

Much like in the case of rigidly rotating models, in both families $f_{WT}^d(I)$ and $f_{PT}^d(I)$, the resulting models are characterized by two scales, associated with the positive constants $A$ and $a$, and two dimensionless parameters $(\Psi, \chi)$, measuring concentration and central rotation strength, respectively. In addition, two new dimensionless parameters, namely $c$ and 
\begin{equation}\label{bbar}
\bar{b}=b\,r_0^{2c}\,a^{-c}~,
\end{equation}
determine the shape of the rotation profile. Variations in the parameter $\bar{b}$ and $c$ are found to be less important. Minor changes in the model properties are found up to  $\bar{b},\,c \approx 4$, above which the precise value of $c$ has only little impact on the properties of the configurations.  

For each family of differentially rotating models, for given values $(\Psi, \bar{b}, c)$ there exists a maximum value of the central rotation strength parameter $\chi_{max}$, corresponding to the last configuration for which the iteration described in Sect.~\ref{Diff2} converges. Such maximally rotating configurations exhibit highly deformed morphologies, characterized by the presence of a sizable central toroidal structure, which will be described in detail in the following sections.   
\begin{figure}[t!]
\centering
\includegraphics[height=.3\textheight]{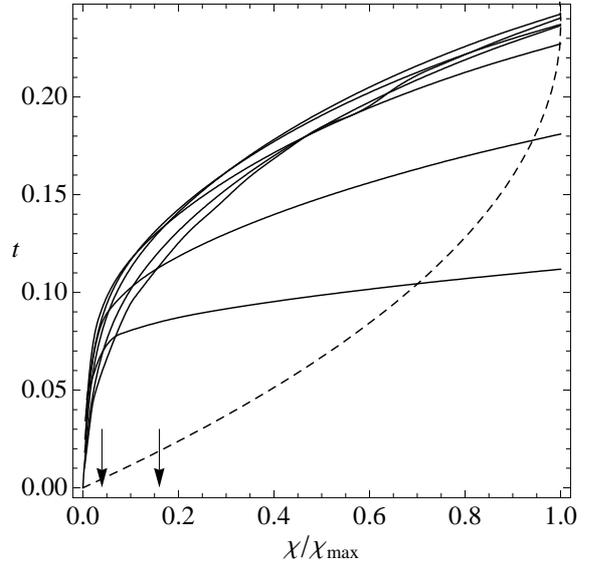}
\caption{ \normalsize Values of the ratio between ordered kinetic energy and gravitational
energy $t=K_{ord}/|W|$ for selected sequences of differentially rotating models 
characterized by $\Psi=1,2,..,7$ (from bottom to top) and $\chi$ in the range $[0,\chi_{max}]$; the remaning parameters are fixed at $\bar{b}=c=1$. The arrows mark the threshold values of $\chi$ for the moderate and rapid rotation regimes, illustrated in Fig.~\ref{parspacediff}. For comparison, the dashed line represents the values of $t$ for the sequence of Maclaurin oblate spheroids (with $e < 0.92995$; this eccentricity value corresponds to spheroids with maximum value of the rotation parameter $\chi_{max}=0.11233$). }  
\label{tchi2}
\end{figure}

As for the parameter space of rigidly rotating models, it is useful to introduce different rotation regimes, defined on the basis of the deviations from spherical symmetry introduced by the presence of differential rotation. With particular reference to the parameter space of the models defined by $f_{WT}^d(I)$, we introduce some threshold values in central dimensionless angular velocity, which, in this family of models, is related to the central rotation strength parameter $\chi$ by the relation $\hat{\omega}=3 \chi^{1/2}$, as for the rigidly rotating models. In particular, configurations in the moderate rotation regime have $\hat{\omega}/\hat{\omega}_{max}<0.2$ (the thin-striped area in Fig.~\ref{parspacediff}), are quasi-spherical in the outer parts, while they are progressively more flattened when approaching the central region, as the value of $\chi$ increases. For the models falling in this rotation regime, the central toroidal structure is absent or, when low values of the concentration parameter $\Psi$ are considered, not significant. Configurations with $0.2< \hat{\omega}/\hat{\omega}_{max}<0.4$ (the wide-striped area in Fig.~\ref{parspacediff}) are defined as rapidly rotating models. The extreme rotation regime is defined by the condition $\hat{\omega}/\hat{\omega}_{max}>0.4$ (the gray area in Fig.~\ref{parspacediff}); in this case, the models always show a central toroidal structure, which becomes more extended as the central rotation strength increases. In particular, in the last regime, the entire volume of a configuration is dominated by the central toroidal structure.

As for the rigidly rotating models described in Sect. 2, a global kinematical characterization is offered by the parameter $t=K_{ord}/|W|$, defined as the ratio between ordered kinetic energy and gravitational energy. Figure \ref{tchi2} illustrates the relation between the two parameters, for models with selected values of $\Psi$, $\bar{b}$, and $c$. Note that the transition from rapid to extreme rotation corresponds to values of the parameter $t$ in the range $[0.075,0.135]$ (the precise value depends on the value of $\Psi$). A naive application of the Ostriker \& Peebles (\cite{OstPee73}) criterion, which states that axisymmetric stellar systems with $t > 0.14$ are dynamically unstable with respect to bar modes, would suggest that the majority of the models in the extreme rotation regime are dynamically unstable. A detailed stability analysis of the configurations in the three rotation regimes has been performed by means of specifically designed N-body simulations and will be presented in a separated paper (Varri et al. \cite{VarAAS}; Varri et al. in preparation). 

To provide a systematic description of the intrinsic and projected properties, we will study the equilibrium configurations as sequences of models characterized by a given value of concentration $\Psi$, in the range $[1,7]$, and increasing values of $\chi$, up to the maximum value allowed; such sequences are constructed by fixing $\bar{b}=c=1$ (unless otherwise stated). 

\section{Intrinsic properties of the differentially rotating models}
\label{DiffIntr}

\subsection{The intrinsic density profile}
\label{DiffIntr1}

The relevant density profile is obtained by integration in velocity space of the distribution function $f_{WT}^d(I)$ (see Eq.~(\ref{fW})). It is convenient to introduce in the velocity space a spherical coordinate system $(v,\mu,\lambda)$, in which $v$ is magnitude of the velocity vector, while $\mu$ and $\lambda$ are the polar and azimuthal angle, respectively. After some manipulation, the density profile can be expressed in dimensionless form as
\begin{eqnarray}\label{rhoW}
\hat{\rho}_{WT}(\hat{r},\theta,\psi)&=&\frac{3}{4} \mathrm{e}^{\psi}\int_0^{\psi} ds\, 
\mathrm{e}^{-s} s^{1/2} \int_{-1}^{+1}dt\;g(s,t,\hat{r},\theta)\nonumber \\
&&-\psi^{3/2}-\frac{2}{5}\psi^{5/2}~,
\end{eqnarray}
where the function in the integrand is defined as
\begin{equation}\label{g}
g(s,t,\hat{r},\theta)= \exp\left( \frac{3 \bar{\omega} \sqrt{2s}\,t\,\hat{r} 
\sin\theta}{1+\bar{b}[\sqrt{2s}\,t\,\hat{r} \sin\theta]^{2c}}\right)~;
\end{equation}
for completeness, we note that the two dimensionless variables in the double integral can be expressed in terms of the previous variables as $t=\cos\mu$ and $s=a v^2/2$. Because the distribution function depends only on the energy $E$ and the z-component of the angular momentum $J_z$, the resulting models are axisymmetric and therefore the density profile depends only on the radius $\hat{r}$ and the polar angle $\theta$. The density depends on the spatial coordinates explicitly and implicitly, through the dimensionless escape energy $\psi(\hat{r},\theta)$; such explicit dependence is the reason why, in this case, the isodensity, equipotential, and isobaric surfaces are not in one-to-one correspondence, at variance with the family of rigidly rotating models.  The presence in Eq.~(\ref{rhoW}) of the terms with fractional powers of $\psi$ is due to the adopted truncation in phase space; in particular, it is directly related to the presence of the terms $\mathrm{e}^{-aE_0}[-1+a(I-E_0)]$ in Eq.~(\ref{fW}). 

The central value of the density profile depends only on the concentration parameter $\Psi$ and is given by $\hat{\rho}_{WT,0}=(2/5)\mathrm{e}^{\Psi}\gamma\left(7/2,\Psi\right)$, consistent with the central value of the density profile obtained in the nonrotating limit $\hat{\rho}_{WT,S}(\Psi)$ (see Eq.~(\ref{rhoS}) in Appendix \ref{AppA}). This result corresponds to the fact that the integral $I(E,J_z)$ (see Eq.~(\ref{I})) reduces to the Jacobi integral for small values of $J_z$, which implies that the rotation is approximately rigid in the central regions of a configuration (see the next subsection for details); therefore, the mean rotation velocity vanishes at the origin and the density distribution reduces to its nonrotating limit. 

The double integral in Eq.~(\ref{rhoW}) requires a numerical integration (see Appendix \ref{AppC} for details). Some insight into the behavior of the density profile can be gained by calculating the relevant asymptotic expansions in the central region and in the outer parts. Around the origin, up to second order in radius, the density profile reduces to
\begin{eqnarray}\label{rhoWcen}
\hat{\rho}_{WT}(\hat{r},\theta,\Psi)&=&\hat{\rho}_{WT,0} + \frac{1}{2}\mathrm{e}^{\Psi}\gamma\left(\frac{5}{2},\Psi\right)\left[ 
9\chi\sin^2{\theta} \right.\nonumber\\
&& \left. +\left.\frac{\partial^2 \psi}{\partial \hat{r}^2}\right|_0\right]\hat{r}^2 +\mathcal{O}(\hat{r}^4)~,
\end{eqnarray}
which depends explicitly on the concentration $\Psi$ and the central rotation strength $\chi$, and implicitly on the parameters $\bar{b}$ and $c$, through the second order derivative of the escape energy evaluated at $\hat{r}=0$. Such derivative can be calculated from the radial functions given in Eqs.~(\ref{eqint0})-(\ref{eqintn})
\begin{eqnarray}\label{d2psi}
\left.\frac{\partial^2 \psi}{\partial \hat{r}^2}\right|_0&=& \psi_0''(0)U_0(\theta)+\psi_2''(0)U_2(\theta)=\nonumber\\
&=&-3+\frac{C_2}{2}\left(\frac{5}{2}\right)^{1/2}\left(-1+3\cos^2\theta\right)~,
\end{eqnarray}
where the quantity
\begin{equation}\label{C2}
C_2\equiv\frac{18}{5 \hat{\rho}_0}\int_0^{+\infty}d \hat{r}' \frac{1}{\hat{r}'} \hat{\rho}_2(\hat{r}')
\end{equation}
depends implicitly on the parameter $\Psi$ through the function $\hat{\rho}_2(\hat{r})$ and $\hat{\rho}_0$. The quantity $C_2$ is negative-definite since the quadrupole radial function of the density $\hat{\rho}_2(\hat{r})$ is negative on the entire domain of definition of the solution of the Poisson equation; the sign of the quadrupolar function is negative because the configurations in our family of models are always oblate. In passing, we also note that the the expansion around $\hat{r}=0$ of the escape energy up to second order in radius is given by
\begin{equation}\label{psicen}
\psi(\hat{r},\theta)=\Psi+\frac{1}{2}\left.\frac{\partial^2 \psi}{\partial \hat{r}^2}\right|_0 \hat{r}^2 +\mathcal{O}(\hat{r}^4)~.
\end{equation}

\begin{figure}[t!]
\centering
\includegraphics[height=.3\textheight]{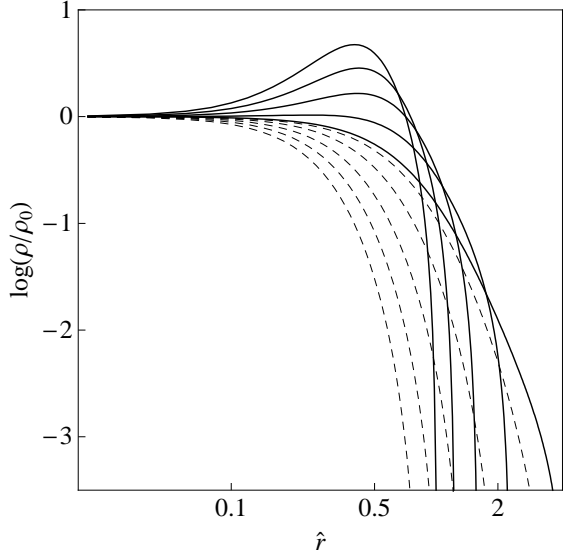}
\caption{ \normalsize Intrinsic density profiles (normalized to the central value) evaluated 
on the equatorial plane (solid lines) and along the $\hat{z}$-axis 
(dashed lines) for a sequence of differentially rotating models defined by $f_{WT}^d(I)$, with 
$\Psi=2$ and $\chi=0.04,0.36,1.00,1.96,3.24$ (from right to left; slower 
rotating models are more extended); the remaining parameters are fixed at $\bar{b}=c=1$.}
\label{dens_diff}
\end{figure}  
 
Since the boundary of a configuration is defined by the condition $\psi(\hat{r},\theta)=0$, the density profile in the outer parts can be evaluated by performing an expansion with respect to $\psi \ll 1$ 
\begin{equation}
\hat{\rho}_{WT}(\hat{r},\theta,\psi)=\frac{9}{5}\chi
\hat{r}^2\sin^2{\theta}\,\psi^{5/2}+\mathcal{O}(\psi^{7/2})~;
\end{equation}
the terms with fractional powers of $\psi$ that appear in Eq.~(\ref{rhoW}) cancel out with the first terms of the expansion of the double integral (which reduces to the incomplete gamma function).    

\begin{figure}[t!]
\centering
\vspace{0.05cm}
\includegraphics[height=.3\textheight]{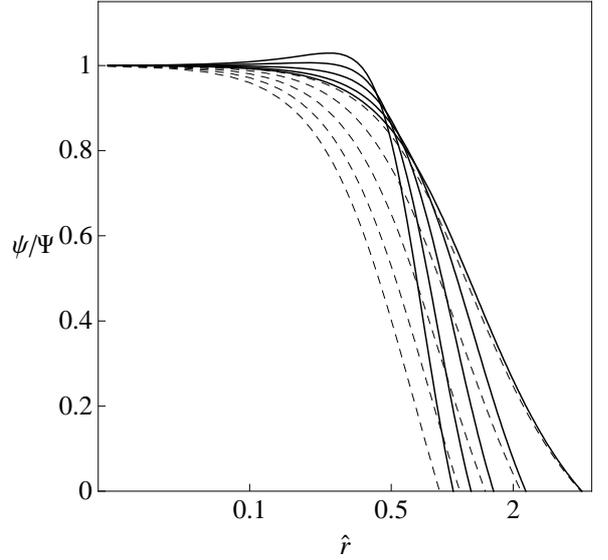}
\caption{ \normalsize Dimensionless escape energy (normalized to the central value) evaluated
on the equatorial plane (solid lines) and along the $\hat{z}$-axis 
(dashed lines) for the sequence of differentially rotating models displayed 
in Fig.~\ref{dens_diff}.}
\label{pot_diff}
\end{figure}

\begin{figure*}
\centering
\includegraphics[height=.235\textheight]{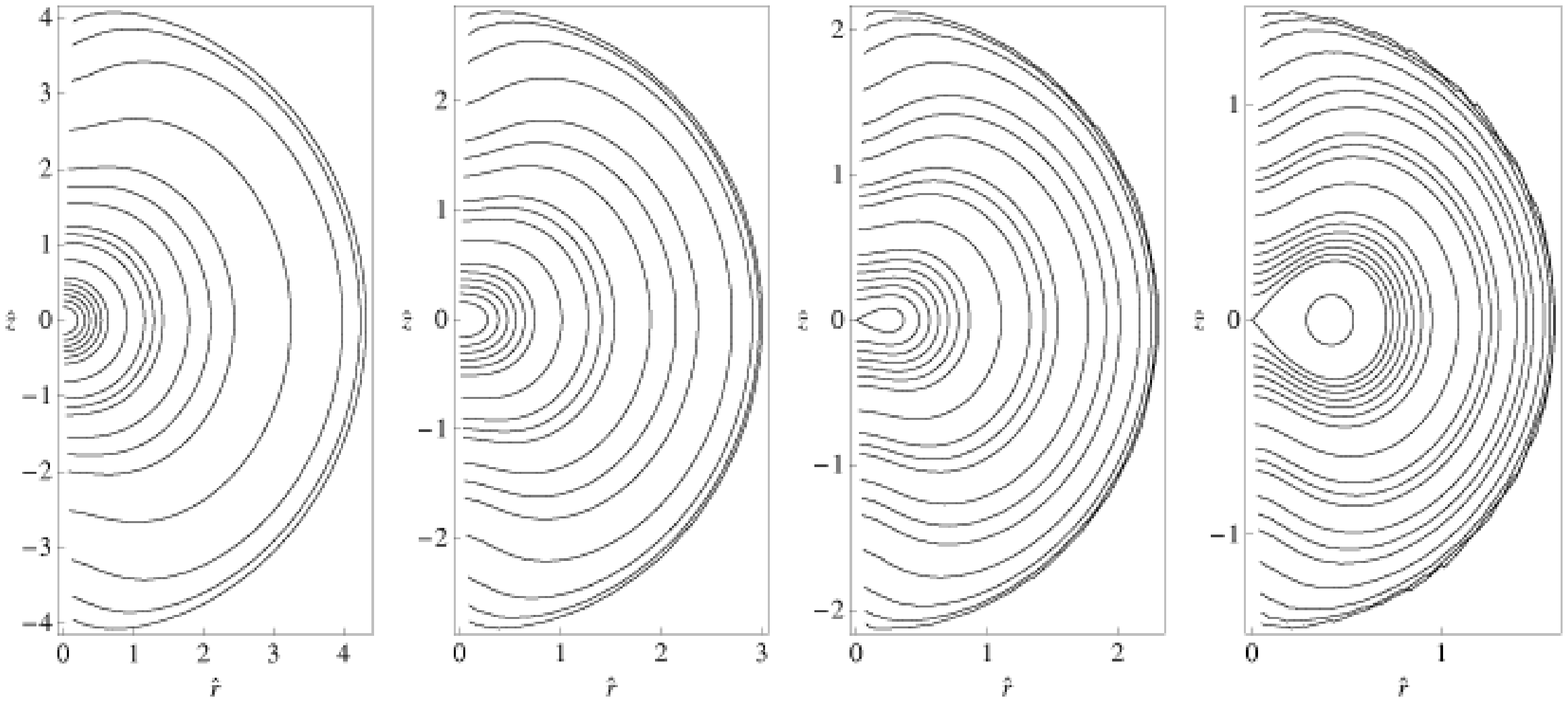}
\includegraphics[height=.235\textheight]{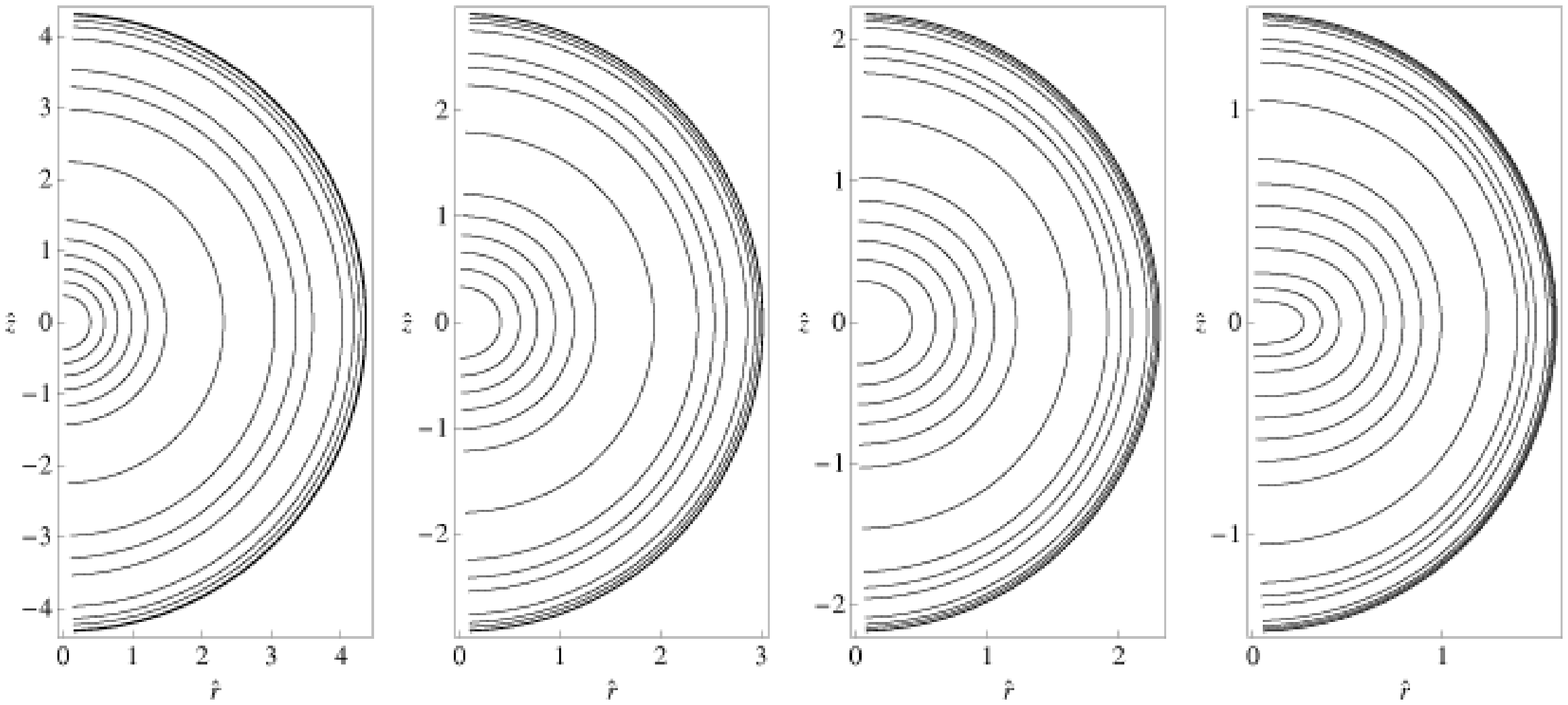}
\includegraphics[height=.235\textheight]{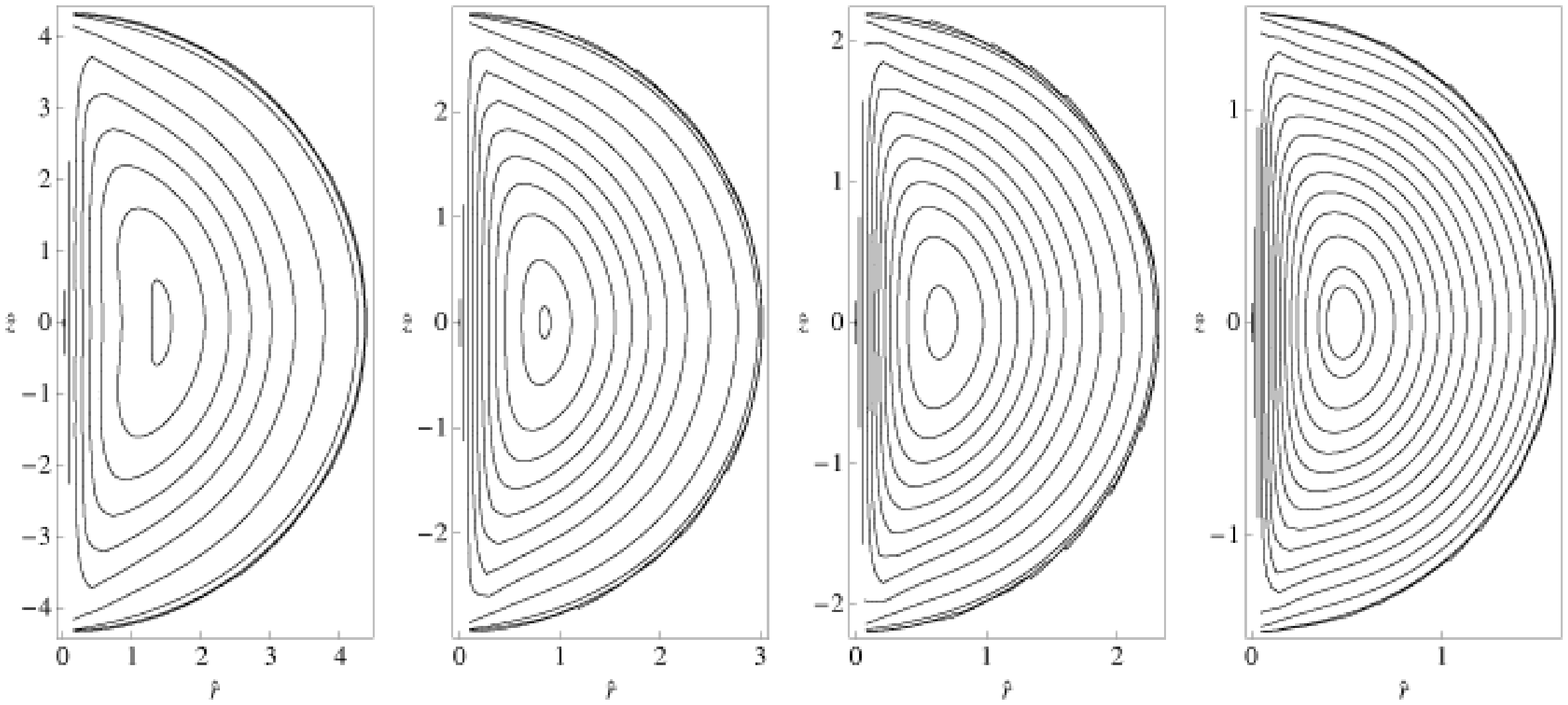}
\includegraphics[height=.235\textheight]{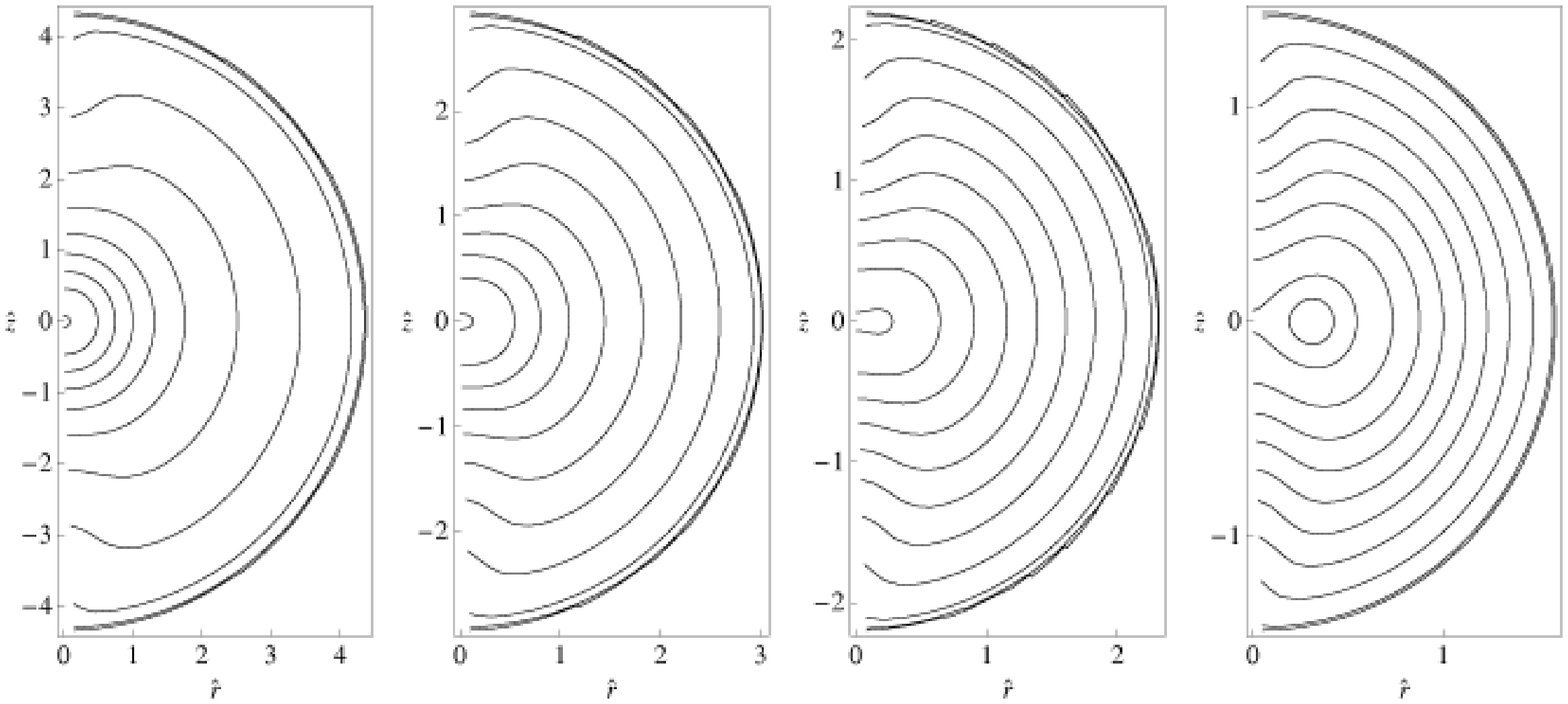}
\caption{ \normalsize Meridional sections of the isodensity, equipotential, 
isovelocity, and isobaric surfaces (from top row to bottom row) for a 
sequence of four differentially rotating models characterized by $\Psi=2$, 
$\bar{b}=c=1$, and $\chi=0.04,0.16,0.36,1.0$ (from left column to right 
column; note the change in the scale of the axes). The first and second models are in the moderate rotation regime, the third has rapid rotation, and the last represents the beginning of the extreme rotation regime. }
\label{intr_maps}
\end{figure*}

The density profiles evaluated on the principal axes for a sequence of models with $\Psi=2$, $\bar{b}=c=1$, and increasing values of the rotation strength parameter $\chi$ are illustrated in Fig.~\ref{dens_diff}; the corresponding dimensionless escape energy profiles are displayed in Fig.~\ref{pot_diff}. Configurations characterized by moderate rotation show monotonically decreasing profiles, whereas models with rapid rotation have the maximum value of the density profile in a position displaced with respect to the origin; in the extreme rotation regime, also the maximum value of the escape energy is off-centered. The sections of the isodensity and equipotential surfaces (presented in the first two rows of Fig.~\ref{intr_maps}) clearly show that the offset of the density peak corresponds to the existence of a curious toroidal structure; the condition for the existence of such structure is discussed in Sect.~\ref{DiffIntr3}.

\subsection{The intrinsic kinematics}
\label{DiffIntr2}

The calculation of the first order moment in velocity space of the distribution function $f_{WT}^d(I)$ confirms that only the azimuthal component of the mean velocity is nonvanishing. The mean velocity in dimensionless form
\begin{eqnarray}\label{uW}
{\langle\hat{v}_{\phi}\rangle}_{WT}(\hat{r},\theta,\psi)&=&\frac{3}{2^{3/2}\hat{\rho}_{WT}} \int_0^{\psi} ds\,
 s \int_{-1}^{+1}dt\;t\;[g(s,t,\hat{r},\theta)\mathrm{e}^{-s}\mathrm{e}^{\psi}\nonumber\\
&& -\ln g(s,t,\hat{r},\theta)]
\end{eqnarray}
can be calculated by numerically. As for the density profile, it is useful to evaluate the asymptotic expansion of Eq.~(\ref{uW}) in the central regions and in the outer part of a given configuration. By performing a first order expansion in the radius with respect to the origin, we found that the mean rotation velocity reduces to the following expression    
\begin{equation}\label{uWcen}
{\langle\hat{v}_{\phi}\rangle}_{WT}(\hat{r},\theta)=3\chi^{1/2}\hat{r}\sin{\theta}+\mathcal{O}(\hat{r}^3)~,
\end{equation}
which corresponds to rigid rotation, with dimensionless angular velocity $\hat{\omega}=3\chi^{1/2}$; the expression does not depend on the concentration parameter, consistent with the asymptotic properties of the integral $I(E,J_z)$.

In the outer parts of the models the mean rotation velocity is of order $\mathcal{O}(\psi)$ and thus vanishes at the boundary. The relevant numerical coefficients depend on the value of the parameter $c$; for example, for $c=1$ we have
\begin{equation}\label{uWbou}
{\langle\hat{v}_{\phi}\rangle}_{WT}(\hat{r},\theta,\psi)=\frac{2(2+6\bar{b}\hat{r}^2 \sin^2 \theta+ 9 \chi \hat{r}^2 \sin^2 \theta)}{21(\chi^{1/2}\hat{r}\sin \theta)}\,\psi~.
\end{equation}

\begin{figure}[t!]
\centering
\includegraphics[height=.3\textheight]{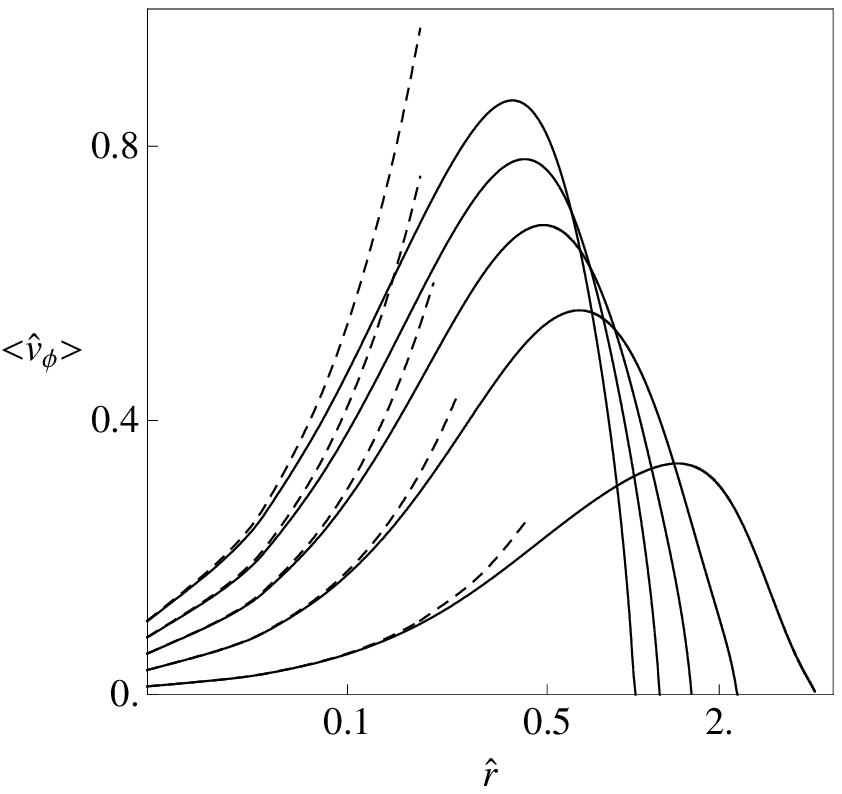}
\caption{ \normalsize Mean rotation velocity profiles on the equatorial plane 
(solid lines) for the sequence of differentially rotating models defined 
by $f_{WT}^d(I)$, with $\Psi=2$ and $\chi=0.04,0.36,1.00,1.96,3.24$ (the same 
sequence displayed in Figs.~\ref{dens_diff} and \ref{pot_diff}, from bottom to top). Dashed lines indicated the 
asymptotic behavior in the central regions, which corresponds to rigid 
rotation (see Eq.~(\ref{uWcen}) and Sect.~\ref{DiffIntr2}).}
\label{vel_diff}
\end{figure}

The mean rotation velocity profiles on the equatorial plane for a selected sequence of models are displayed in Fig.~\ref{vel_diff}: as the value of the central rotation strength parameter increases, the mean velocity profile becomes steeper in the inner parts and the maximum value increases; the central part of the profiles is well approximated by rigid rotation, as in Eq.~(\ref{uWcen}).     

By evaluating the second-order moments in velocity space of the distribution function $f_{WT}^d(I)$, it can be easily shown that, in our coordinate system, the pressure tensor $p_{ij}=(\hat{A}/a)\hat{p}_{ij}$ is diagonal and that the radial component is equal to the polar one $\hat{p}_{rr}=\hat{p}_{\theta\theta}$. Therefore, in the following, only the nontrivial components are discussed. The radial and azimuthal components in dimensionless form are given by 
\begin{eqnarray}\label{prrW}
\hat{p}_{W,rr}(\hat{r},\theta,\psi)&=&\frac{3}{4} \mathrm{e}^{\psi}\int_0^{\psi} ds\,
\mathrm{e}^{-s} s^{3/2} \int_{-1}^{+1}dt\;(1-t^2)\;g(s,t,\hat{r},\theta)\nonumber \\
&&-\frac{2}{5}\psi^{5/2}-\frac{4}{35}\psi^{7/2}~,
\end{eqnarray}
\begin{eqnarray}\label{pffW}
\hat{p}_{W,\phi\phi}(\hat{r},\theta,\psi)&=&\frac{3}{2} \mathrm{e}^{\psi}\int_0^{\psi} ds\,
\mathrm{e}^{-s} s^{3/2} \int_{-1}^{+1}dt\;t^2\;g(s,t,\hat{r},\theta)-\nonumber \\ 
&&-\frac{2}{5}\psi^{5/2}-\frac{4}{35}\psi^{7/2}-{\hat{\rho}}_{WT}{\langle\hat{v}_{\phi}}\rangle_{WT}^2~,
\end{eqnarray}
where the presence of the terms with fractional powers of $\psi$ should be interpreted as in Eq.~(\ref{rhoW}). The expansion up to second order in radius  gives
\begin{eqnarray}\label{prrWcen}
\hat{p}_{W,rr}(\hat{r},\theta,\Psi)&=&\hat{\rho}_{WT,0} + \frac{1}{5}\mathrm{e}^{\Psi}\gamma\left(\frac{7}{2},\Psi\right)\left[ 
9\chi\sin^2{\theta} \right.\nonumber\\
&& \left. +\left.\frac{\partial^2 \psi}{\partial \hat{r}^2}\right|_0\right]\hat{r}^2 +\mathcal{O}(\hat{r}^4)~.
\end{eqnarray}
We also found that, to second order in radius, the pressure is isotropic $\hat{p}_{W,\phi\phi}(\hat{r},\theta,\Psi)= \hat{p}_{W,rr}(\hat{r},\theta,\Psi)$. The components of the pressure tensor evaluated at the origin reduces to $\hat{p}_{WT,0}=(4/35)\mathrm{e}^{\Psi}\gamma\left(9/2,\Psi\right)$,
consistent with the value obtained in the nonrotating limit $\hat{p}_{WT,S}(\Psi)$ (see Eq.~(\ref{pS}) in Appendix~\ref{AppA}).
\begin{figure}[t!]
\centering
\includegraphics[height=.3\textheight]{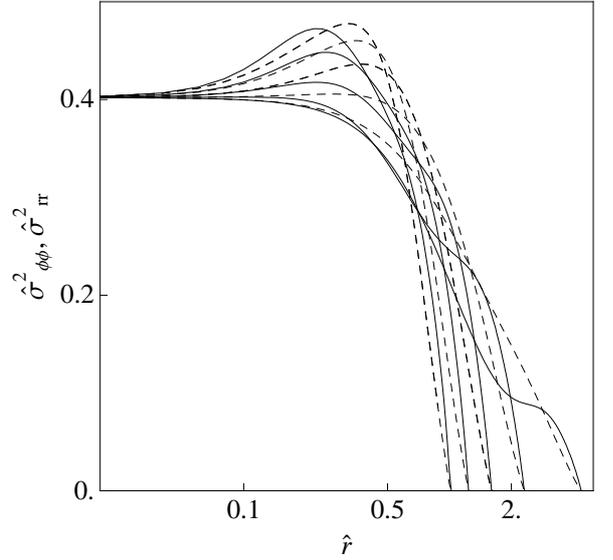}
\caption{\normalsize Squared velocity dispersion profiles for the azimuthal (solid lines) and 
radial component (dashed lines) of the sequence of differentially rotating 
models displayed in Fig.~\ref{vel_diff} (from right to left). The profiles 
are evaluated on the equatorial plane.}
\label{sig_diff}
\end{figure}

\begin{figure}
\centering
\includegraphics[height=.3\textheight]{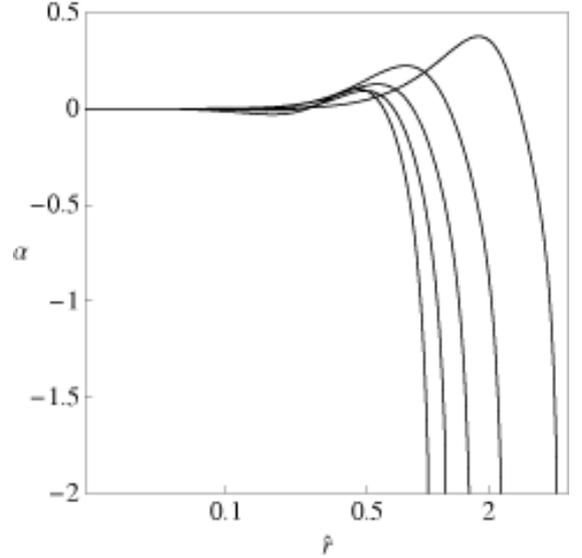}
\caption{\normalsize Radial profiles of the anisotropy parameter for the sequence of 
differentially rotating models displayed in Fig.~\ref{sig_diff}, evaluated 
on the equatorial plane (from right to left). All models are characterized 
by $\alpha=0$ (isotropy) at the center and $\alpha=-2$ (tangentially-biased anisotropy) at the boundary.}
\label{beta_diff}
\end{figure}

The asymptotic behavior of the pressure tensor components in the outer part can be written as
\begin{equation}\label{prrWbou}
\hat{p}_{W,rr}(\hat{r},\theta,\psi)=\frac{18}{35}\chi
\hat{r}^2\sin^2{\theta}\,\psi^{7/2}+\mathcal{O}(\psi^{9/2})~,
\end{equation}
\begin{equation}\label{pffWbou}
\hat{p}_{W,\phi\phi}(\hat{r},\theta,\psi)=\frac{54}{35}
\chi\hat{r}^2\sin^2{\theta}\,\psi^{7/2} +\mathcal{O}(\psi^{9/2})~,
\end{equation}
respectively.

The relation between the dimensionless pressure and velocity dispersion tensor is given by $\hat{\sigma}_{ij}^2=\hat{p}_{ij}/\hat{\rho}$. The profiles of the radial and azimuthal component of the velocity dispersion tensor of a selected sequence of models, evaluated on the equatorial plane, are displayed in Fig.~\ref{sig_diff}. For configurations in the moderate rotation regime the profiles are monotonically decreasing with the radius, with some variations in the slope in the intermediate and external parts, while for configurations in the rapid and extreme rotation regimes, their peak is off-centered. The sections in the meridional plane of the isobaric surfaces, defined with respect to the trace of the pressure tensor, show the presence of a central toroidal structure for the fast rotating models of the sequence (see the last row in Fig.~\ref{intr_maps}).   

\begin{figure*}[t!]
\centering
\includegraphics[height=.41\textheight]{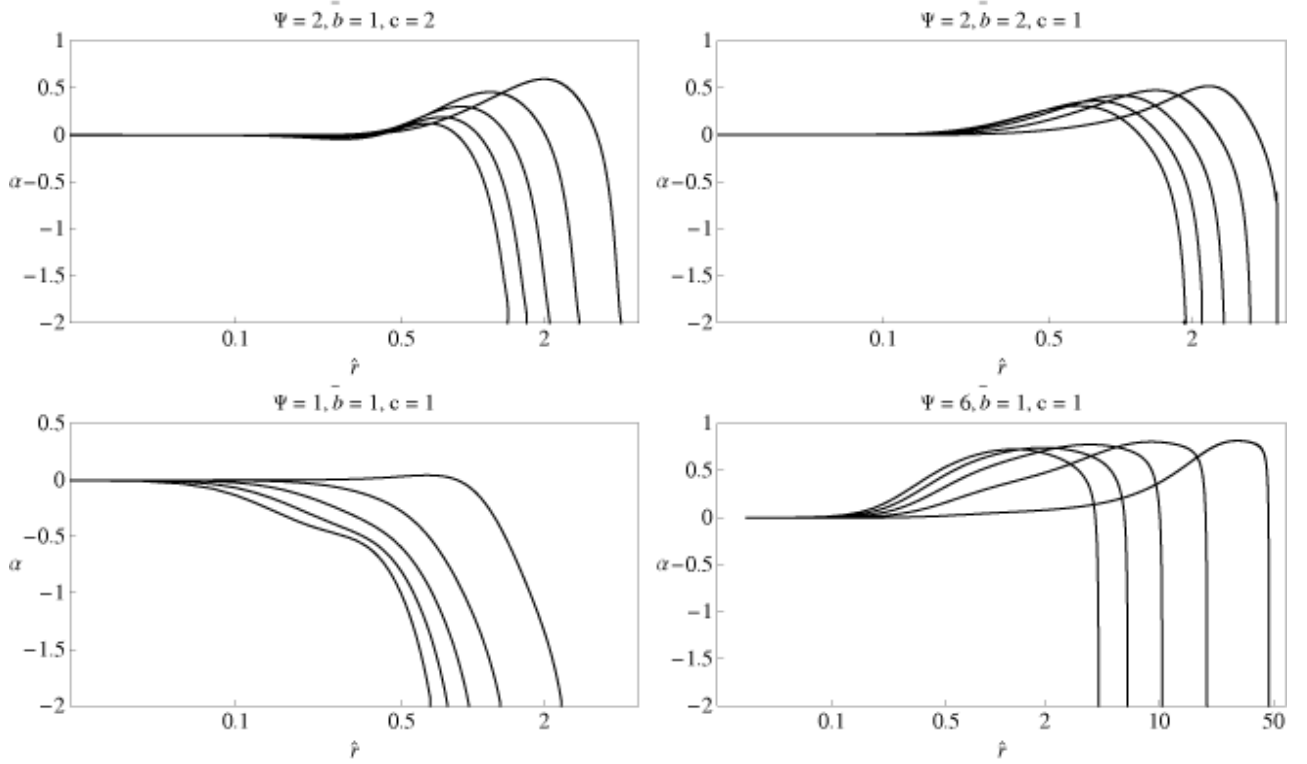}
\caption{\normalsize Radial profiles of the anisotropy parameter for selected sequences 
of differentially rotating models, evaluated on the equatorial plane. 
Top panels: the sequences are characterized by $\Psi=2$, 
$\chi=0.04,0.16,0.36,0.64,1.00$ (in each panel, from right to left); the left and right panels show the effect 
of increasing of parameters $c$ and $\bar{b}$, respectively.
Bottom panels: the sequences are characterized by $\bar{b}=c=1$, $\chi=0.04,0.16,0.36,0.64,1.00$ 
(in each panel, from right to left); the left and right panels show the 
effect of decreasing and increasing the concentration parameter $\Psi$, respectively.} 
\label{alphaall}
\end{figure*}

The intrinsic kinematics can be further characterized by means of the anisotropy parameter\footnote{In the literature, the anisotropy parameter is often defined as $\beta= 1-(\sigma^2_{\phi \phi}+\sigma^2_{\theta \theta})/2 \sigma^2_{rr}$; for axisymmeric systems, for which $\sigma^2_{\theta \theta}=\sigma^2_{rr}$, the relation with the parameter adopted in the present paper is given by $\alpha= 2 \beta$. Here we prefer to keep the same notation used in van Albada \cite{Alb82}, Zocchi et al. \cite{ZBV11}, and in other articles.}
, defined as 
\begin{equation}\label{alpha}
\alpha=1-\frac{p_{\phi\phi}}{p_{rr}}=1-\frac{\sigma_{\phi\phi}^2}{\sigma_{rr}^2}~. 
\end{equation}
From the asymptotic behavior of the pressure tensor components in the central region (see Eq.~(\ref{prrWcen})), we find that $\alpha \rightarrow 0$ for $\hat{r}\rightarrow 0$, while from the expansion in outer parts (see Eqs.~(\ref{prrWbou})-(\ref{pffWbou})), we find that $\alpha \rightarrow -2$ as we approach the boundary. These limiting values do not depend on the dimensionless parameters that characterize the family $f_{WT}^d(I)$. In other words, the central region of the configurations is always characterized by isotropy in velocity space, while the regions next to the boundary show a strong tangentially-biased pressure anisotropy. The radial profiles of the anisotropy parameter for a selected sequence of models are displayed in Fig.~\ref{beta_diff}. The values of $\alpha$ in the intermediate region of a given configuration depend on the values of the relevant dimensionless parameters $(\Psi,\bar{b},c)$. In fact, from an exploration of the entire four-dimensional parameter space, we found that, by increasing the value of $c$, the portion of a model dominated by radial anisotropy becomes slightly less extended, while, if the value of $\bar{b}$ is increased, it increases (see Fig.~\ref{alphaall}, top panels). In turn, by increasing the value of the concentration, as measured by $\Psi$, the models are characterized by a significant radially-biased pressure anisotropy, which appears also in the intermediate region. By decreasing the value of the concentration, the intermediate region turns out to be dominated by tangential anisotropy, like in the outer parts (Fig.~\ref{alphaall}, bottom panels).            

\subsection{The condition for the existence of the central toroidal structure}
\label{DiffIntr3}

In general, configurations with rapid rotation may exhibit highly deformed morphologies. In the context of rotating fluids, the regime of strong differential rotation has been successfully explored, at least for polytropes. Stoeckly (\cite{Sto65}) and Geroyannis (\cite{Ger90}) found that rapidly rotating polytropic fluids show a central toroidal structure; in addition, a self-consistent method for the construction of rapidly differentially rotating fluid systems with a great variety of shapes has been proposed by Hachisu (\cite{Hac86}). In stellar dynamics, this regime has been rarely explored: Lynden-Bell (\cite{Lyn62}) and  Prendergast \& Tomer (\cite{PreTom70}) noted that some models with strong differential rotation show the density peak in a ring on their plane of symmetry.   
\begin{figure}[t!]
\centering
\includegraphics[height=.3\textheight]{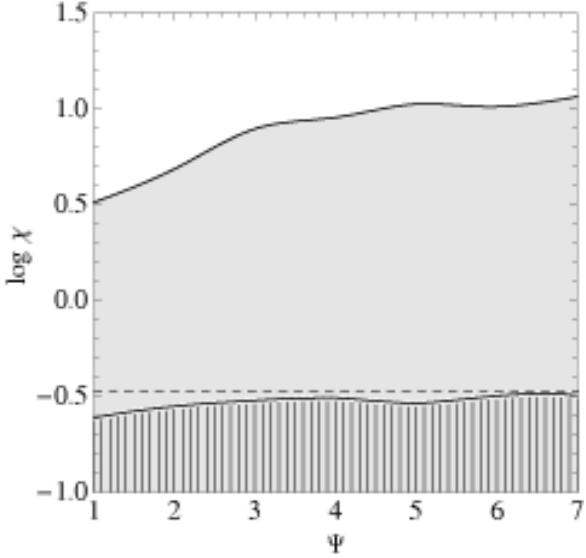}
\caption{\normalsize Two-dimensional parameter space of differentially rotating models $(\Psi, \chi)$; the remaining 
parameters are fixed at values $\bar{b}=c=1$. The upper solid line marks 
the maximum admitted values of $\chi$, as in Fig. \ref{parspacediff}. The lower solid 
line marks the values of $\chi$ at which the models start to exhibit a central 
toroidal structure; the dashed line marks the upper limit of the condition 
for the existence of the central toroidal structure (see Eq.~(\ref{condtor})).}
\label{parspacetor}
\end{figure}

The existence of a central toroidal structure in a given model of our family can be studied by using the asymptotic expansion of the density in the central regions, expressed in Eq.~(\ref{rhoWcen}). In fact, if the second order derivative of the density with respect to the radius is positive, then the maximum value of the radial density profile is displaced from the geometric center of the configuration, so that a central toroidal structure is formed. Therefore, the relevant condition on the sign of the derivative can be expressed as
\begin{equation}\label{condexp}
9 \chi \sin^2\theta -3 +\frac{C_2}{2} \left(\frac{5}{2}\right)^{1/2}\left(-1+3\cos^2\theta\right) > 0~,    
\end{equation}   
which, on the equatorial plane (i.e., at $\theta=\pi/2$), reduces to a simple condition for the central rotation strength parameter
\begin{equation}\label{condtor}
\chi > \frac{1}{3} +\frac{C_2}{18}\left(\frac{5}{2}\right)^{1/2}~.    
\end{equation}   
The expression on the right-hand side depends implicitly on the dimensionless parameters $\Psi$, $\bar{b}$, and $c$ through the quantity $C_2$, defined by Eq.~(\ref{C2}). Since $C_2$ is negative-definite, the condition $\chi> 1/3$ provides the upper limit to Eq.~(\ref{condtor}). Actually, the models characterized by values of $\chi$ immediately above the threshold given by Eq.~(\ref{condtor}), show a very small central toroidal structure, with a shallow increase of the density with respect to the geometric center of the configuration  (e.g., for the model with $\Psi=2$ and $\chi=0.36$, illustrated in Fig.~\ref{dens_diff}, the density peak, at the center of the toroidal structure, is merely  $\log(\rho/\rho_0)=0.014$ and the central structure itself is barely visible in the meridional sections of the isodensity surfaces, depicted in the third panel of the first row of Fig. \ref{intr_maps}). To construct a model with a sizable central toroidal structure, the value of the parameter $\chi$ should be at least twice the threshold given by the above-mentioned condition (e.g., see the last panel of the first row of Fig. \ref{intr_maps}). In conclusion, by comparing Figs. \ref{parspacediff} and \ref{parspacetor}, we note that, for configurations in the moderate rotation regime (i.e., with $\hat{\omega}/\hat{\omega}_{max} \la 0.2$), the central toroidal structure is absent or very small, while starting from the rapid rotation regime, the structure is always present and becomes more extended as the value of the central rotation strength increases.     
\begin{figure}[t!]
\centering
\includegraphics[height=.3\textheight]{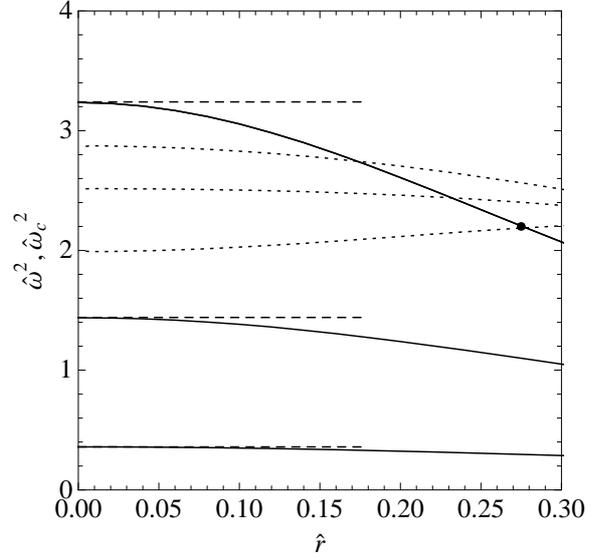}
\caption{\normalsize Squared angular velocity (solid lines) evaluated in the central parts of 
the equatorial plane for the first three differentially rotating models 
displayed in Fig.~\ref{intr_maps}, that is with $\Psi=2$, $\bar{b}=c=1$, and 
$\chi=0.04,0.16,0.36$ (from bottom to top). Dashed lines indicate the 
asymptotic behavior at small radii, characterized by solid-body 
rotation, that is constant angular velocity. 
Dotted lines represent the angular velocity associated with the circular orbit of a single star,
evaluated on the equatorial plane for the same models (from top to bottom). 
For the first two models  $\hat{\omega}_c > \hat{\omega}$, while for the 
third $\hat{\omega}>\hat{\omega}_c$ in the inner part, where the toroidal 
structure is present (the black filled circle marks the position where $\hat{\omega}=\hat{\omega}_c$)~.}
\label{circ}
\end{figure}

\begin{figure*}
\centering
\includegraphics[height=.173\textheight]{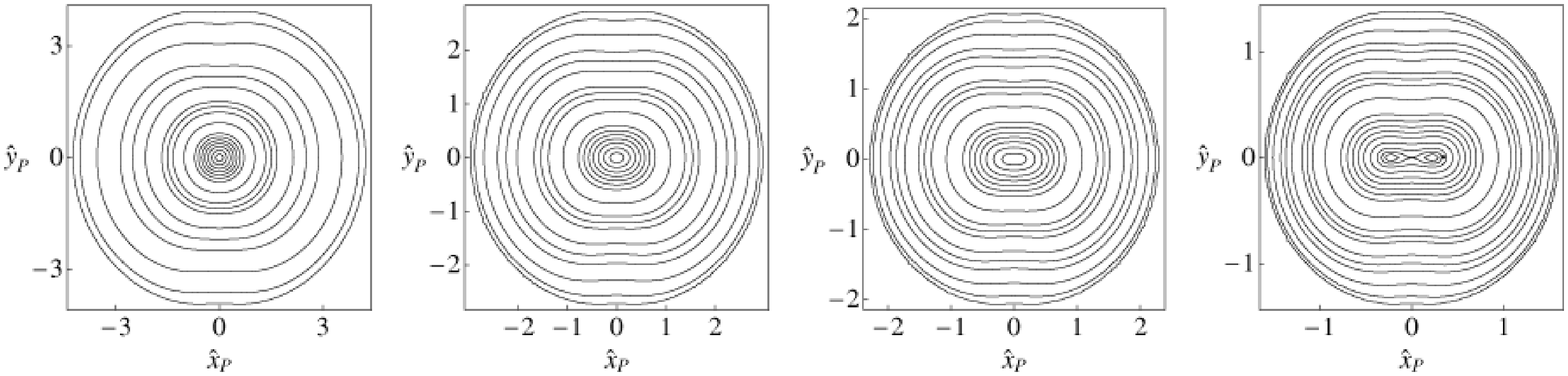}
\includegraphics[height=.173\textheight]{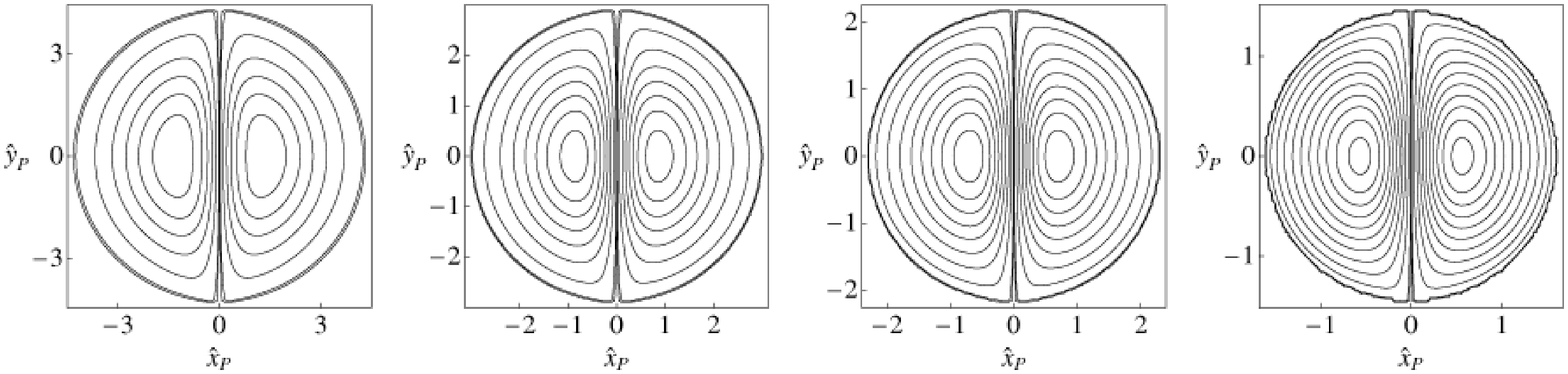}
\includegraphics[height=.173\textheight]{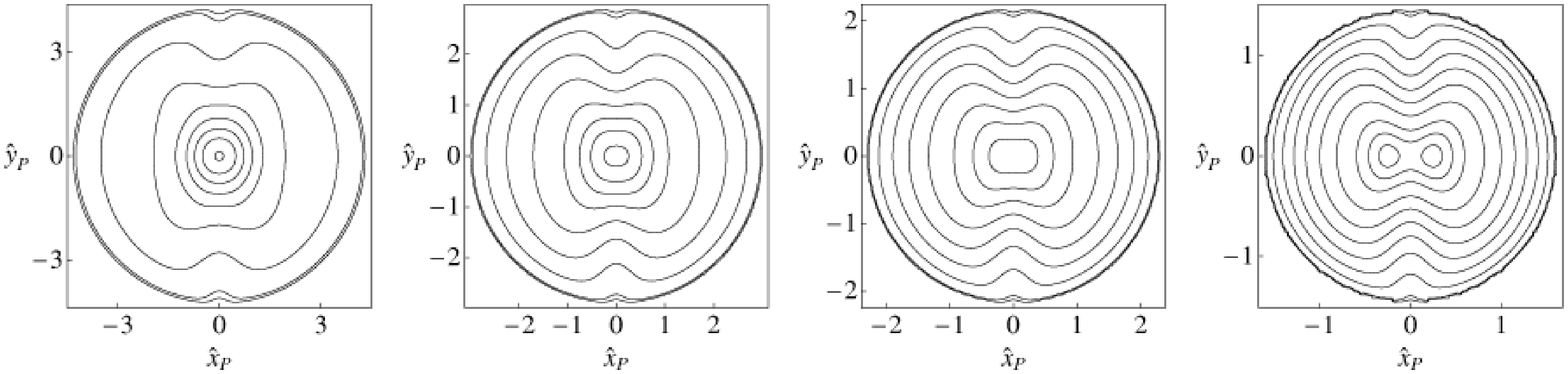}
\caption{\normalsize Contour maps of the surface density, mean line-of-sight velocity, and 
line-of-sight velocity dispersion (from top row to bottom row) of the 
differentially rotating models with $\Psi=2$, $\bar{b}=c=1$, and 
$\chi=0.04,0.16,0.36,1.00$ (from left to right, as in Fig.~\ref{intr_maps}), projected 
along the $\hat{x}$-axis of the intrinsic coordinate system (``edge-on'' view, so that $\hat{x}_P$ and $\hat{y}_P$ correspond to the $\hat{y}$ and $\hat{z}$ axes of the intrinsic system, respectively).
In the panels in the first row, solid lines represent the isophotes corresponding 
to selected values of $\Sigma/\Sigma_0$ in the range $[1.02, 10^{-7}]$; only 
the last (fastest rotating) model shows values $\Sigma/\Sigma_0>1$, when the toroidal structure appears. Panels in the second row illustrate the contours 
of the dimensionless rotation velocity in the range $[0.5,10^{-5}]$ at intervals of $0.05$ 
(from left to right, the values of the innermost contours are $0.25,0.35,0.4,0.45$, 
respectively.). Panels in the last row show the contours of the 
projected velocity dispersion in the range $[0.4,10^{-5}]$ at intervals 
of $0.05$ (the values of the innermost contours are $0.3$ for the first 
three models and $0.4$ for the last one). }
\label{proj_maps}
\end{figure*}
The presence of the central toroidal structure can be also be interpreted in terms of to the intrinsic kinematical properties the models. From the radial component of the Jeans equation, expressed in dimensionless spherical coordinates,
\begin{equation}\label{jeans}
\frac{1}{\hat{\rho}}\frac{\partial\hat{\rho}}{\partial \hat{r}}\frac{\hat{\sigma}_{rr}^2}{\hat{r}}=\frac{1}{\hat{r}}\frac{\partial \psi}{\partial \hat{r}} + \frac{\langle \hat{v}_{\phi}\rangle^2}{\hat{r}^2}-\left[\frac{1}{\hat{r}}\frac{\partial \hat{\sigma}_{rr}^2}{\partial \hat{r}}+\frac{\hat{\sigma}_{rr}^2-\hat{\sigma}_{\phi\phi}^2}{\hat{r}^2}\right]~,
\end{equation}
the sign of the first order derivative of the density with respect to the radius (on the left-hand side of Eq.~(\ref{jeans})) depends on (i) the difference between the angular velocity associated with the circular orbit of a single star $\hat{\omega}_c=[-(1/\hat{r})\partial_{\hat{r}}\psi]^{1/2}$ and the angular velocity of the model $\hat{\omega}$, associated to the mean rotation velocity $\langle \hat{v}_{\phi}\rangle=\hat{\omega}\hat{r}\sin\theta$, (ii) a more complex pressure term (in square brackets on the right-hand side of Eq.~(\ref{jeans})). To check for the presence of the central toroidal structure, it is sufficient to study the Jeans equation in the central region of the model. Therefore, by inserting the relevant asymptotic expansions for the mean velocity and the escape energy (see Eqs.~(\ref{uWcen}) and (\ref{psicen})), the term (i) reduces to the expression indicated on the left-hand side of Eq.~(\ref{condexp}), 
and, by inserting the relevant asymptotic expansions for the pressure tensor components, the term (ii) can be written as
\begin{eqnarray}
&&\left[9 \chi \sin^2\theta -3 +\frac{C_2}{2} \left(\frac{5}{2}\right)^{1/2}\left(-1+3\cos^2\theta\right)\right]\times \nonumber\\
&&\left[1-\frac{5}{7}\frac{\gamma(9/2,\Psi)\gamma(5/2,\Psi)}{\gamma(7/2,\Psi)^2} \right].
\end{eqnarray}  
By combining the asymptotic expressions of terms (i) and (ii), it follows\footnote{The coefficient given by $1-(5/7)[\gamma(9/2,\Psi)\gamma(5/2,\Psi)]/\gamma(7/2,\Psi)^2$ is nonnegative for every value of $\Psi$.} that the requirement of a positive density gradient on the left-hand side of the Jeans equation is equivalent to the condition expressed by Eq.~(\ref{condexp}).  

In conclusion, we have independently tested the validity of the condition for the existence of the central toroidal structure derived at the beginning of this section. We have also shown that the requirement of positivity of term (i) in the radial Jeans equation is a {\em necessary and sufficient} condition for the presence of the central toroidal structure. In other words, the central toroidal structure exists if and only if, in the central region, the angular velocity associated with the internal rotation is higher than the angular velocity associated with the circular orbit of a single star.  Figure~\ref{circ} shows the relevant angular velocities for the first three models of the sequence considered in Fig. \ref{intr_maps}; it is apparent that, for the configuration in which the central toroidal structure is present, $\hat{\omega}> \hat{\omega}_c$.    
This result strictly depends on the adopted truncation in phase space for the distribution function $f_{WT}^d(I)$; in fact, in Appendix \ref{AppB} we show that for the alternative distribution function $f_{PT}^d(I)$ , the condition on the angular velocities is a necessary but {\em not} sufficient condition for the existence of the central toroidal structure.         

\begin{figure}[t!]
\centering
\vspace{0.23cm}
\includegraphics[height=.30\textheight]{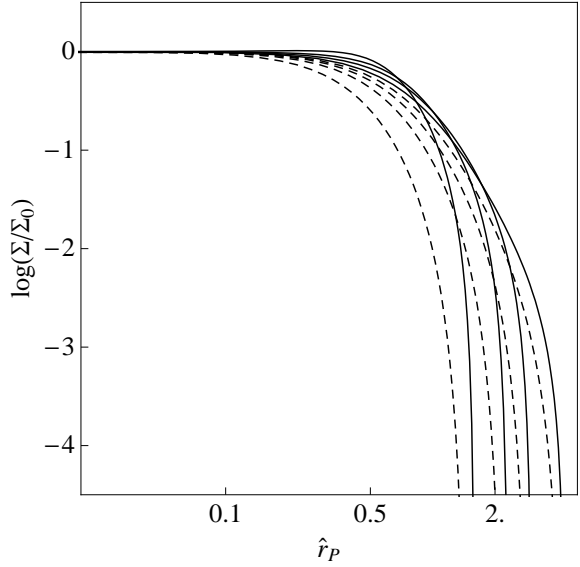}
\caption{\normalsize Surface density profiles (normalized to the central value) of the 
differentially rotating models with $\Psi=2$, $\bar{b}=c=1$, and 
$\chi=0.04,0.16,0.36,1.00$ (from right to left; same sequence illustrated in Fig.~\ref{proj_maps}), evaluated 
along the $\hat{x}_P$-axis (solid lines) and $\hat{y}_P$-axis (dashed lines) 
of the projection plane (``edge-on'' view).}
\label{projdens}
\end{figure}
\subsection{The condition for maximally rotating models}
\label{DiffIntr4}

If we consider a sequence of models with fixed values of $\Psi, \bar{b}, c$ and increasing values of $\chi$, the central toroidal structure becomes progressively more extended and characterized by a larger aperture angle (i.e., the angle spanned at small radii by the isodensity surface that represents the boundary of the central toroidal structure). For high values of the central rotation strength parameter $\chi$, a smaller central toroidal structure appears also in the equipotential surfaces (see the escape energy profile of the fastest rotating model in Fig.~\ref{pot_diff}). These two structures can be characterized in terms of $\theta_p$ and $\theta_d$, defined as the complements of the semi-aperture angle of the inner cusp of toroidal structures in the equipotential and isodensity surface, respectively. In fact, since the boundary of the central cusp of the toroidal structures, is defined by $\psi(\hat{r},\theta)=\Psi$, and $\hat{\rho}(\hat{r},\theta,\psi)=\hat{\rho}_0$, by using the relevant asymptotic expansion up to second order in $\hat{r}$ given in Eqs.~(\ref{psicen}) and (\ref{rhoWcen}), the following expressions for the angles are obtained:  
\begin{equation}\label{thetap}
\cos^2 \theta_p = \frac{3+(5/2)^{1/2} C_2/2}{(3/2)(5/2)^{1/2} C_2}~,
\end{equation}
\begin{equation}\label{thetad}
\cos^2 \theta_d = \frac{3+(5/2)^{1/2} C_2/2-9\chi}{(3/2)(5/2)^{1/2} C_2-9\chi}~.
\end{equation}
The two angles are related, because 
\begin{equation}\label{thetas}
\cos^2 \theta_d = \frac{(3/2)(5/2)^{1/2} C_2 \cos^2 \theta_p -9\chi}{(3/2)(5/2)^{1/2} C_2-9\chi}~.
\end{equation}
Note that $\theta_p$ and $\theta_d$ decrease as the values of $|C_2|$ and $\chi$ increase (we recall that $C_2$ is negative-definite), that is the corresponding semi-aperture angles become larger. 

For high values of the central rotation strength $\chi$, which imply a high degree of quadrupolar deformation, as measured by the quantity $|C_2|$, it is easy to see that $\cos^2\theta_p \rightarrow 1/3$; interestingly, we also found (numerically) that a limiting value exists also for $\theta_d$, given by $\cos^2 \theta_d \rightarrow 2/3$.  By inserting the limiting value of $\theta_p$ in Eq.~(\ref{thetas}), the condition for the maximum value of the angle $\theta_d$ (i.e., $\cos^2 \theta_d < 2/3$) can be translated in a simple condition for the central rotation strength parameter
\begin{equation}
\chi < \frac{1}{6}\left(\frac{5}{2}\right)^{1/2}|C_2|~,
\end{equation} 
which basically provides a limit to the deformation induced by the central rotation itself. The previous condition has been checked numerically and we found that, for values of $\chi$ above the threshold, the iteration scheme used for the solution of the Poisson equation does not converge. 

\section{Projected properties of the differentially rotating models}
\label{DiffProj}
\subsection{The surface density profile}
\label{DiffProj1}
The calculation of the projected properties has been performed by following the same projection rules described in Sect.~\ref{Unif4}, that is the line of sight corresponds to the $\hat{z}_P$-axis of a new frame of reference in which the projection plane is denoted by $(\hat{x}_P,\hat{y}_P)$. In particular, we studied in detail the ``edge-on'' view, that is the projection along the $\hat{x}$-axis of the intrinsic frame of reference ($\hat{z}_P=\hat{x}$, $\hat{x}_P=\hat{y}$, and $\hat{y}_P=\hat{z}$, i.e., the viewing angles of the rotation matrix are $\theta=\pi/2$, $\phi=0$). The dimensionless surface density distribution $\hat{\Sigma}(\hat{x}_P,\hat{y}_P)$ has been calculated by numerical integration of Eq.~(\ref{Sigma}) on an equally-spaced square cartesian grid on the projection plane.
\begin{figure}[t!]
\centering
\includegraphics[height=.31\textheight]{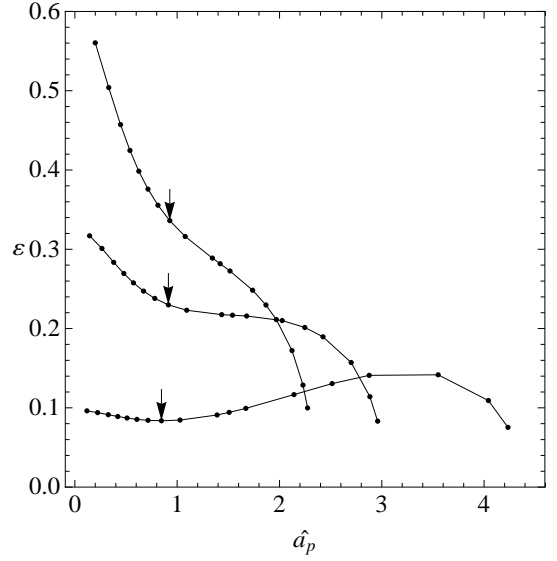}
\caption{\normalsize Ellipticity profiles, as functions of the semi-major axis of the 
  projected image $\hat{a}_P$, of the first three differentially rotating 
  models illustrated in Fig.~\ref{projdens} (from bottom to top; ``edge-on'' view). Dots 
  correspond to the isophotes shown in the first row of 
  Fig.~\ref{proj_maps} and arrows mark the position of the half-light isophote.}
\label{projellip}
\end{figure}

The first row of Fig.~\ref{proj_maps} shows the contour maps of the surface density of selected differentially rotating models in the moderate and rapid rotation regime. As result of projection, the dimples on the rotation axis, which are prominent in the corresponding intrinsic isodensity surfaces (see Fig.~\ref{intr_maps} for the meridional sections of the same sequence of models), are less pronounced. In addition, the central toroidal structure in the projected density distribution is visible only if it has a reasonable size in the intrinsic density distribution (see third and last model of the sequence illustrated in Fig.~\ref{proj_maps}).

The surface density profiles of the same sequence of differentially rotating models, evaluated along the principal axes of the projection plane, are presented in Fig.~\ref{projdens}. For the configurations in which the central toroidal structure in projection is absent, we calculated the relevant ellipticity profiles, as functions of the semi-major axis of the projected image $\hat{a}_P$, and they are reported in Fig.~\ref{projellip}. As expected, the configurations in the moderate rotation regime are characterized by nonmonotonic ellipticity profiles, while models in the rapid rotation regime have monotonically {\em decreasing} profiles; to some extent, this morphological feature is complementary to that of the uniformly rotating models, in which the configurations are always characterized by monotonically {\em increasing} ellipticity profiles. Interestingly, the behavior of the ellipticity profiles does not necessarily correlate with the mean line-of-sight velocity profiles (see next subsection), that is configurations with a nonmonotonic velocity profile may have a monotonic ellipticity profile.        

In addition, the isophotes of models show clear departures from a pure ellipse, which, at variance with the family of rigidly rotating models, can be characterized as a ``boxy'' overall trend, particularly evident in the intermediate parts of the configurations (see third and fourth panels of the first row of Fig.~\ref{proj_maps}).    
\begin{figure}
\centering
\includegraphics[height=.3\textheight]{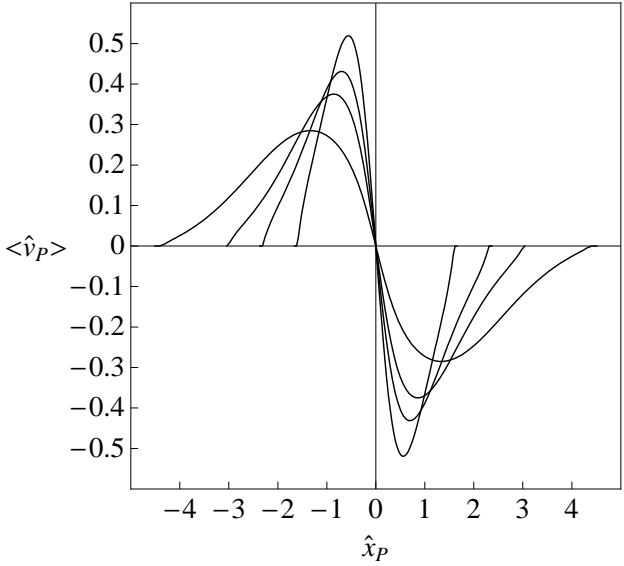}
\caption{\normalsize Mean line-of-sight rotation velocity profiles of the 
differentially rotating models with $\Psi=2$, $\bar{b}=c=1$, and 
$\chi=0.04,0.16,0.36,1.00$ (slower rotating models are more extended; same 
sequence illustrated in Fig.~\ref{projdens}), evaluated 
along the $\hat{x}_P$-axis of the projection plane (``edge-on'' view).}
\label{projvel}
\end{figure}

\subsection{The line-of-sight velocity distribution}
\label{DiffProj2}
The projected velocity moments are calculated by integrating along the line of sight (weighted by the intrinsic density) the corresponding intrinsic quantities. As for the surface density distribution, we studied in detail the kinematics projected along the $\hat{x}$-axis of the intrinsic coordinate system. Therefore, the  dimensionless mean line-of-sight velocity and the line-of-sight velocity dispersion can be written as
\begin{equation}\label{vP}
\langle \hat{v}_P\rangle (\hat{x}_P,\hat{y}_P)= -\frac{1}{\hat{\Sigma}(\hat{x}_P,\hat{y}_P)} 
\int_{\hat{z}_{sp}}^{\hat{z}_{sp}} d \hat{z}_P\,\hat{\rho}(\hat{\bf r}_P)\, 
\frac{\langle \hat{v}_{\phi}\rangle\,\hat{x}_P}{(\hat{z}_P^2+\hat{x}_P^2)^{1/2}}~,
\end{equation}
\begin{eqnarray}\label{sigmaP}
&&\hat{\sigma}_P^2(\hat{x}_P,\hat{y}_P)=\frac{1}{\hat{\Sigma}(\hat{x}_P,\hat{y}_P)} \int_{\hat{z}_{sp}}^{\hat{z}_{sp}} d \hat{z}_P\,
\hat{\rho}(\hat{\bf r}_P)\,\left[ \frac{1}{\hat{z}_P^2+\hat{x}_P^2}(\hat{z}_P^2 \hat{\sigma}_{rr}^2+
\right.\nonumber \\
&& \left.\hat{x}_P^2\hat{\sigma}_{\phi\phi}^2+\langle \hat{v}_{\phi}\rangle^2 \hat{y}^2 )+ \langle \hat{v}_P\rangle^2+2\langle \hat{v}_P\rangle\langle \hat{v}_{\phi}\rangle\frac{\hat{x}_P}{(\hat{z}_P^2+\hat{x}_P^2)^{1/2}} \right]~.
\end{eqnarray}

Selected mean line-of-sight velocity profiles (for the sequence of models presented in Fig.~\ref{projdens}), evaluated along the $\hat{x}_P$-axis of the projection plane, are illustrated in Fig.~\ref{projvel}, where the sign of the mean velocity in the two half-planes is consistent with Eq.~(\ref{vP}). As the value of the central rotation strength parameter increases, the slope of the inner part of the profile becomes steeper, since the asymptotic behavior of the intrinsic velocity in the central regions is approximately that of a rigid rotation, with the angular velocity proportional to $\chi^{1/2}$ (see Eq. ~(\ref{uWcen})); in addition, because the entire configuration becomes more compact, the radial position of the peak of the velocity profile shrinks progressively.
\begin{figure}
\centering
\includegraphics[height=.3\textheight]{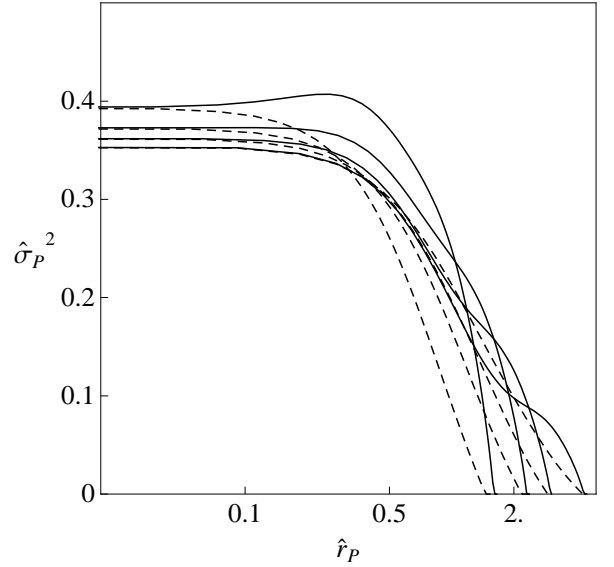}
\caption{\normalsize Squared line-of-sight velocity dispersion profiles of the 
differentially rotating models illustrated in Fig.~\ref{projvel}, evaluated 
along the $\hat{x}_P$-axis (solid lines) and $\hat{y}_P$-axis (dashed lines) 
of the projection plane (``edge-on'' view).}
\label{projdisp}
\end{figure}
\begin{figure}
\centering
\includegraphics[height=.3\textheight]{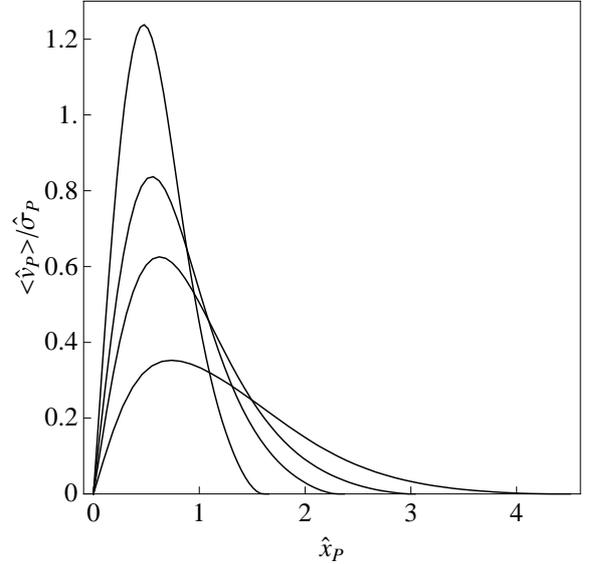}
\caption{\normalsize Ratio of mean line-of-sight (rotation) velocity to the line-of-sight velocity dispersion of the 
differentially rotating models illustrated in Figs.~\ref{projvel} and \ref{projdisp}, evaluated 
along the $\hat{x}_P$-axis of the projection plane (``edge-one'' view). The first and second models are in the moderate rotation regime, the third has rapid rotation, and the last represents the beginning of the extreme rotation regime.}
\label{projrat}
\end{figure}

The line-of-sight velocity dispersion profiles of the same sequence of models, evaluated along the principal axes of the projection plane, are presented in Fig.~\ref{projdisp}. For configurations in the moderate rotation regime, the variations in the slope at intermediate and outer radii, which characterize the azimuthal component of the intrinsic velocity dispersion tensor, are still visible in projection, while the inner part of the profile has a flat core. For models in the rapid and extreme rotation regime, the maximum value of the profile is displaced from the geometric center; in this case, the inner part of the profile has a nontrivial gradient.         
\begin{figure*}
\centering
\includegraphics[height=.29\textheight]{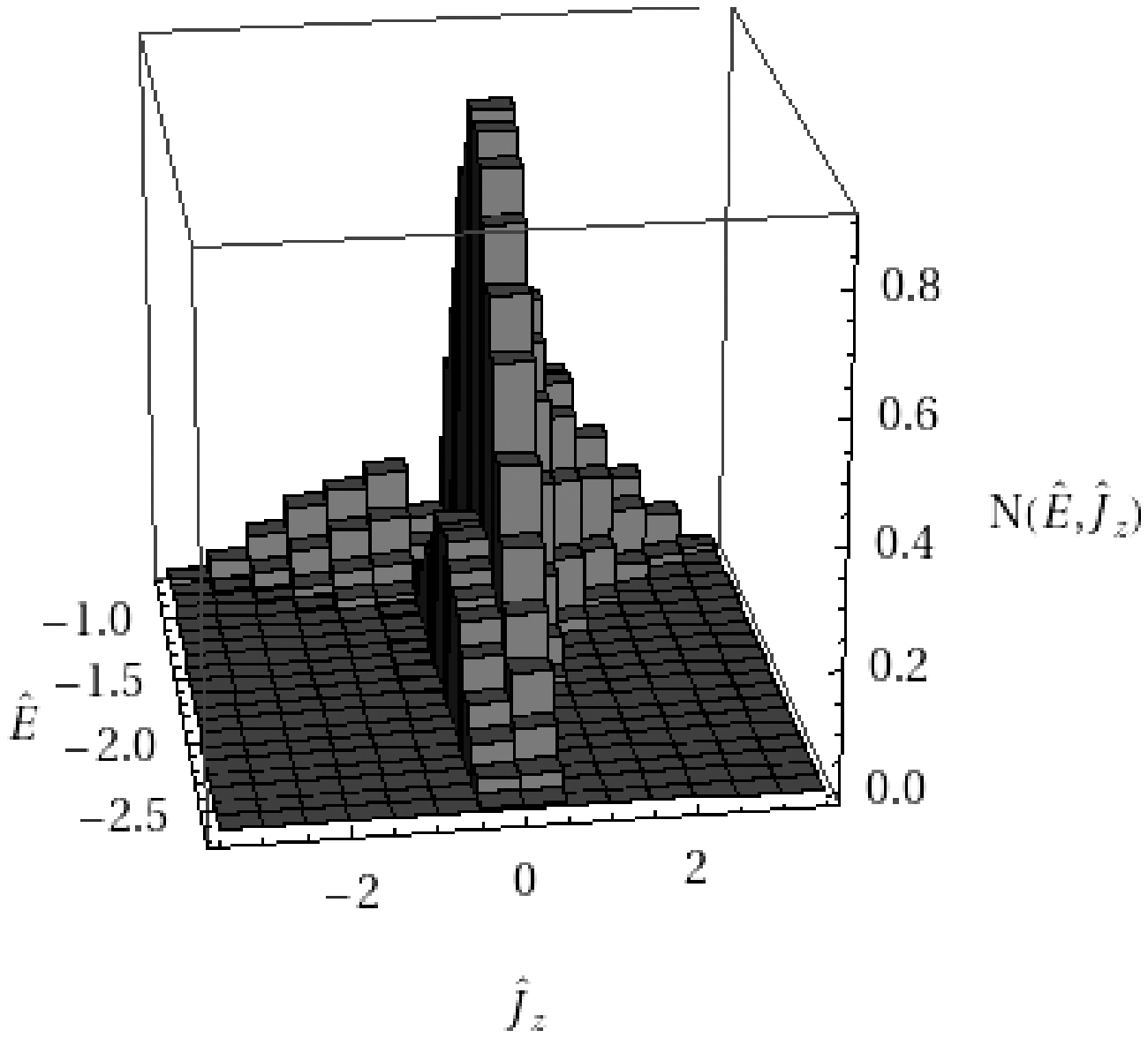}
\includegraphics[height=.28\textheight]{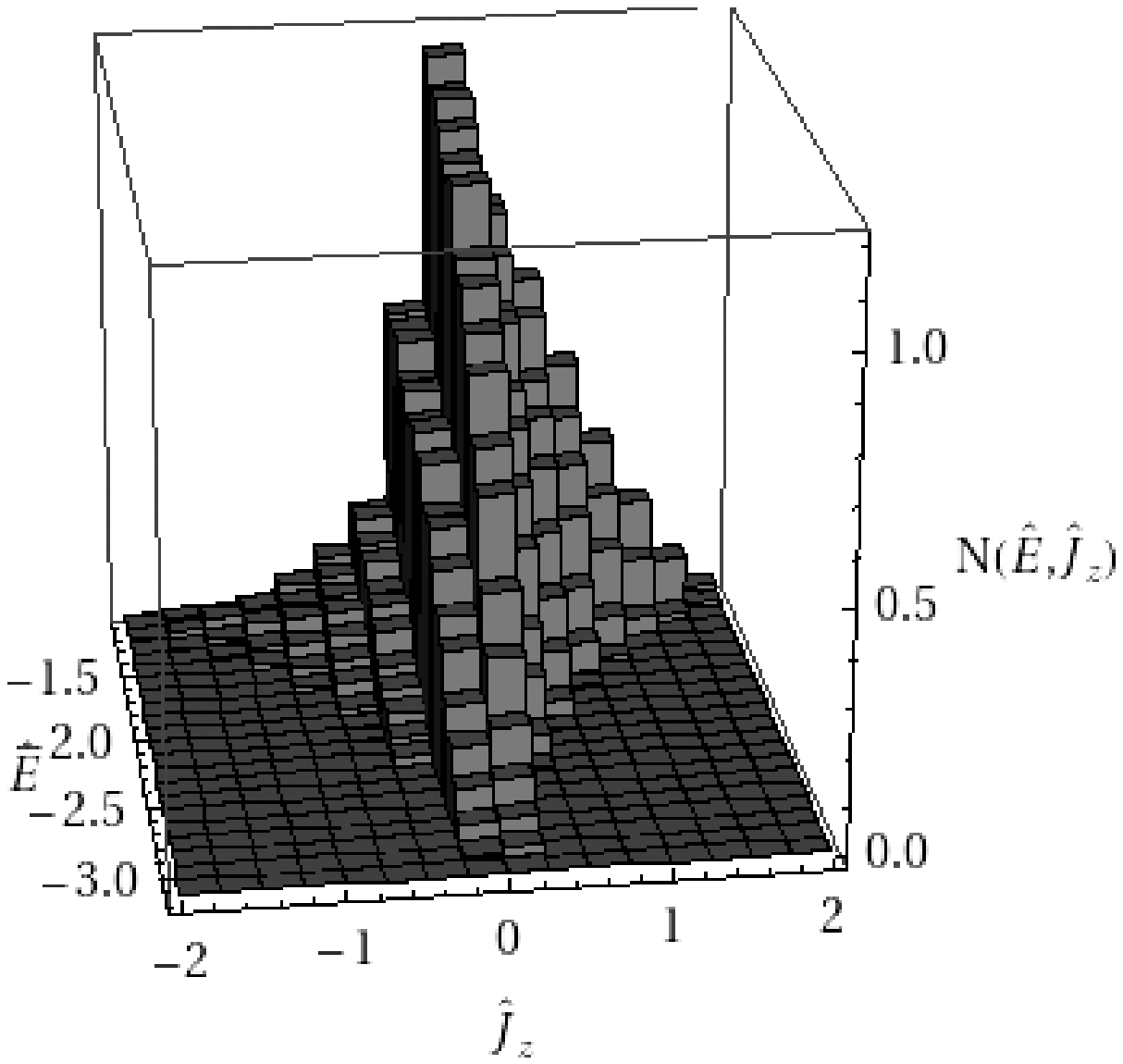}
\vspace{0.3cm}\\	
\includegraphics[height=.22\textheight]{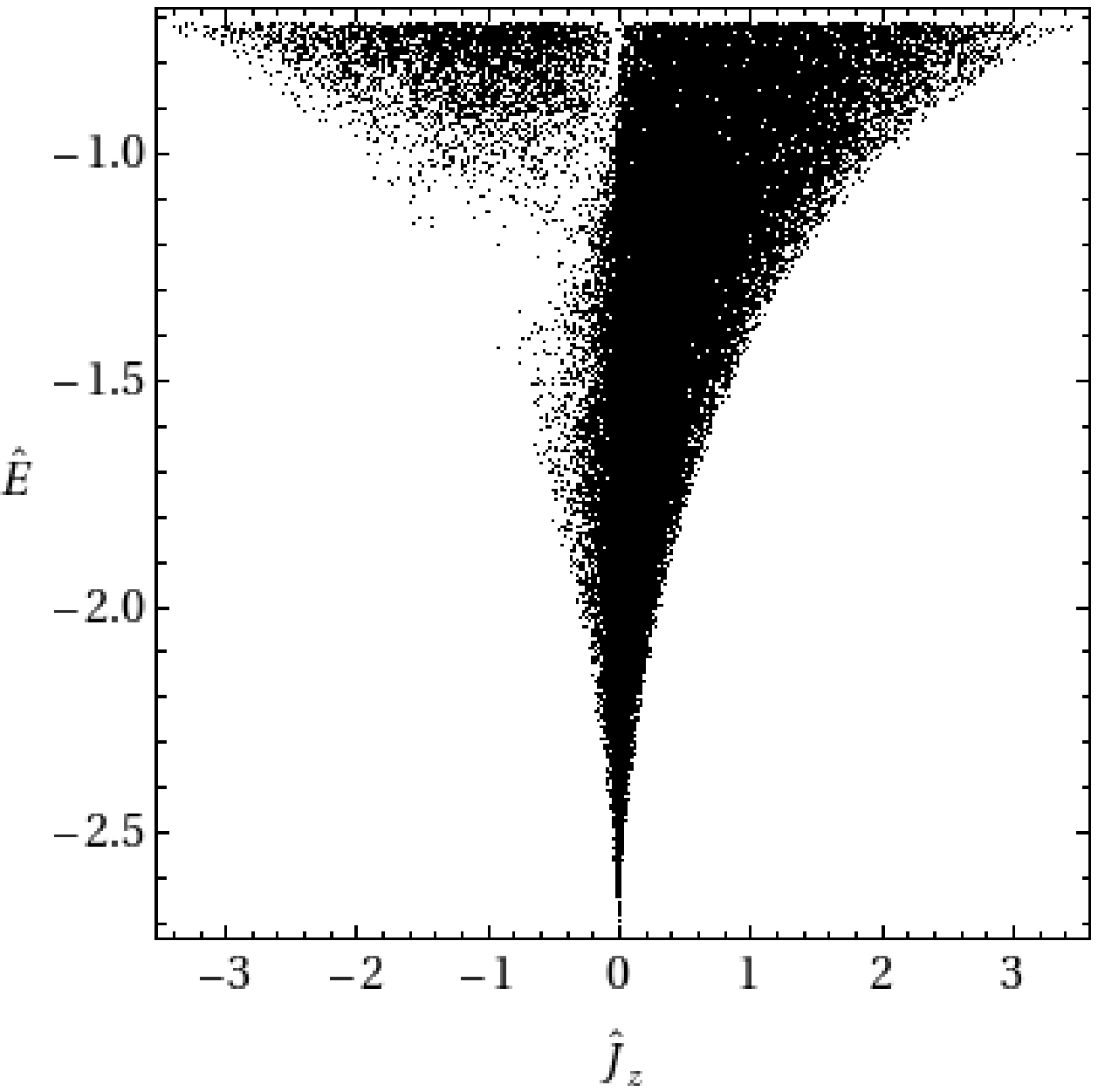}
\hspace{2.5cm}
\includegraphics[height=.22\textheight]{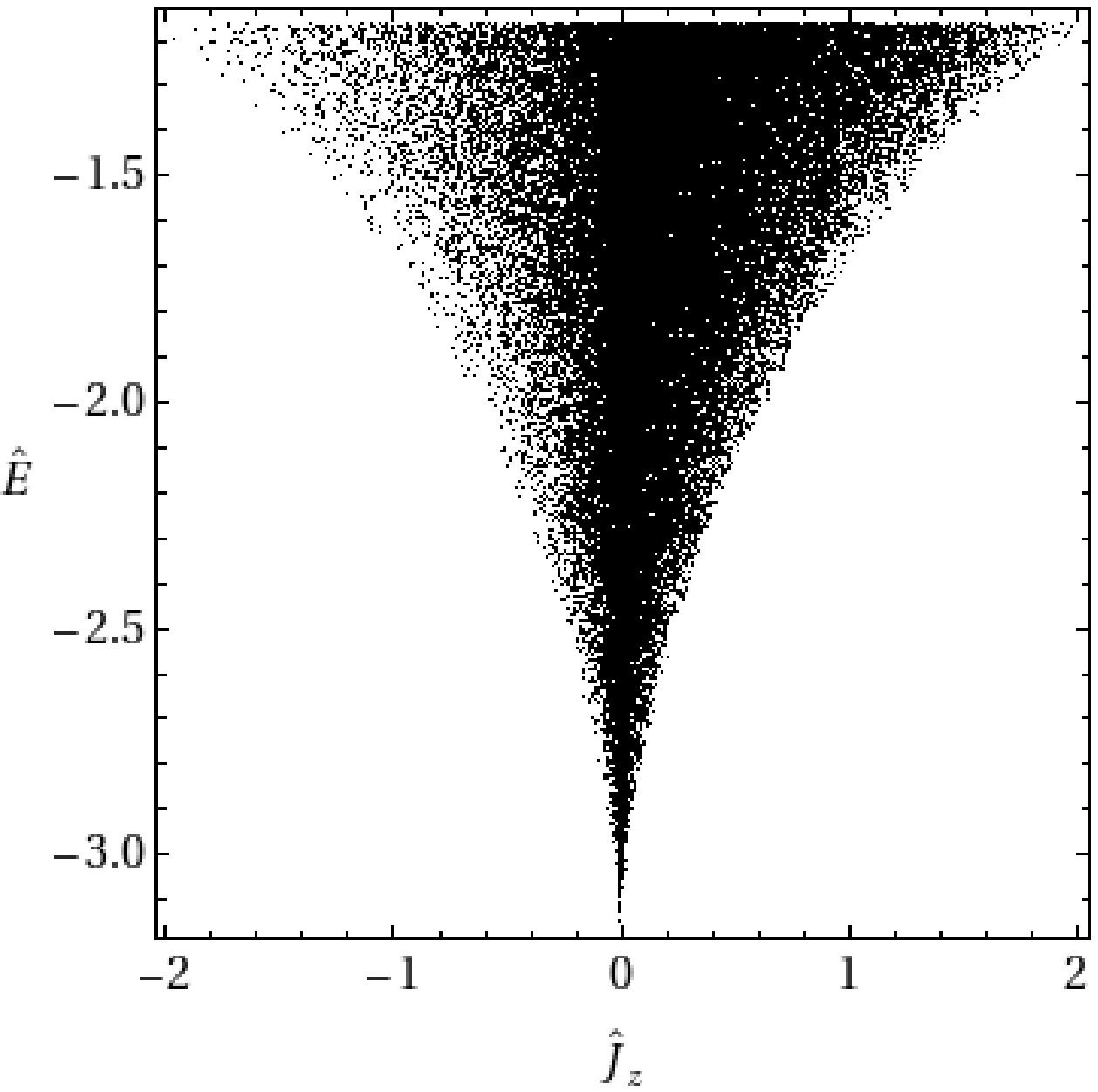}
\hspace{1.3cm}
\caption{\normalsize Top panels: histogram (scaled to unity) of the phase space density $N(\hat{E},\hat{J}_z)$ for a differentially rotating model characterized by Wilson truncation (left) with $\Psi=2,\chi=0.16,\bar{b}=c=1$, and for one characterized by plain truncation (right) with $\Psi=2,\chi=0.36,\bar{b}=c=1$. Both models are characterized by $\hat{\omega}/\hat{\omega}_{max}\approx0.2$. Bottom panels: Lindblad diagrams for the same models presented in the top panels. The graphs have been obtained by a Monte Carlo sampling of the relevant distribution functions with $65\,536$ particles.   }
\label{trunc}
\end{figure*}
We also calculated the ratio of the mean line-of-sight (rotation) velocity to the line-of-sight velocity dispersion, as a measure of the amount of ordered motions compared to random motions (see Fig.~\ref{projrat} for relevant profiles evaluated along the $\hat{x}_P$-axis, for the sequence of models discussed above). The models in the moderate rotation regime show values that are consistent with those observed in Galactic globular clusters (see Introduction). The configurations in the rapid and extreme rotation regime are characterized by higher values of the ratio, even greater than one; such high values are measured only in the class of elliptical galaxies known as fast rotators (see Davies et al. \cite{Dav83}). Of course, the values of this ratio strongly depend on the line of sight on which the projection is performed (we recall that here we illustrate the results obtained only from the ``edge-on'' view, which is the most favorable for the detection of ordered motions).   
\section{Effect of truncation in phase space}
\label{Trun}
To explore the nontrivial effects of the two options for the truncation prescription of the family of differentially rotating models (see Sect. \ref{Diff1}; see also Hunter \cite{Hun77}), we selected two representative models and studied the relevant phase space density $N(\hat{E},\hat{J}_z)$, defined in such a way that the total dimensionless mass is given by $\hat{M}=\int N(\hat{E},\hat{J}_z)d\hat{E}\,d\hat{J}_z$, with the specific energy $\hat{E}=a\,E$ and z-component of the specific angular momentum $\hat{J_z}=(\hat{\bf r}\times \hat{\bf v})_z=\hat{v}_{\phi}\hat{r}\sin \theta$. The histograms of the phase space density, constructed by means of a Monte Carlo sampling of the distribution functions, are presented in the top panels of Fig. \ref{trunc}. 

We also constructed the corresponding Lindblad diagrams (Lindblad 1933), that is the representation in the plane $(\hat{E},\hat{J}_z)$ of orbits in phase space in a given model (see bottom panels of Fig. \ref{trunc}). For any given $\hat{J}_z$, the orbit with the lowest energy corresponds to a circular orbit in the equatorial plane; therefore, in both families of models, the circular orbits form the lower cuspy boundary of the model in the diagram. The upper boundary corresponds to the energy truncation, introduced by the cut-off constant $\hat{E}_0$. Of course, since both families of models are characterized by internal rotation, the distribution of the orbits is asymmetric with respect to the z-component of the angular momentum, but different truncation prescriptions determine a different distribution of the orbits in phase space.  In particular, the Wilson truncation introduces a sharp depopulation of retrograde orbits with intermediate energy, while the plain truncation is associated with a smooth decrease of the number of retrograde orbits with high energy. Alternative options for the truncation prescription in rotating models correspond to a distribution of orbits in different regions of the diagram (see Fig.~1 in Rowley \cite{Row88}). Note that a truncation with respect to the Jacobi integral $H= E-\omega J_z$ only, as in our family of rigidly rotating models $f_K^r(H)$ (see Eq.~(\ref{fK})), corresponds to a region in the diagram bounded by a straight line (with $\omega$ as slope) and the curves corresponding to the energy of circular orbits. 

In addition, as the concentration of the models increases, the cusp of the lower boundary, given by the energy of circular orbits, becomes sharper; as noted also by Rowley (\cite{Row88}), this is the reason why centrally concentrated configurations with rigid rotation cannot be very flat, except for the outer parts, as described in Sect. \ref{Unif4}.       

\section{Discussion and conclusions}
\label{Concl}

In this paper we have constructed two new families of self-consistent axisymmetric models of quasi-relaxed stellar systems, characterized by the presence of internal rotation (see Table 1); a full description in terms of the intrinsic and projected properties has been provided. The main results can be summarized as follows:
\begin{itemize}
\item Driven by general statistical mechanics considerations, we started by constructing a family of rigidly rotating dynamical models; this family is defined as an extension of the King (\cite{Kin66}) models to the case of axisymmetric equilibria flattened by solid-body rotation, with the relevant distribution function dependent only on the Jacobi integral. The configurations have been constructed self-consistently by solving the Poisson-Laplace equation for the mean-field potential by means of a perturbation method (described in Paper I), using a measure of the rotation strength as the expansion parameter. The two-dimensional parameter space which characterizes the family (concentration and rotation strength) can be described in terms of two regimes. Models in the low-deformation regime are almost indistinguishable from the corresponding spherical King models. Highly-deformed models are quasi-spherical in the central regions and show significant deviations from spherical symmetry in the outer parts; in particular, they are flattened toward the equatorial plane and exhibit a sort of ``disky'' appearance. The resulting eccentricity profile is a monotonically increasing function of radius; the  (finite) central value can be expressed analytically in terms of the rotation strength parameter. From the kinematical point of view, the models are characterized by pressure isotropy and cylindrical rotation.         
\item In view of possible applications to globular clusters, we have constructed a second family of dynamical models, characterized by differential rotation, designed to be approximately rigid in the central regions and to vanish in the outer parts, where the imposed energy truncation is effective. In this case, the relevant Poisson equation is solved by means of a spectral iteration method based on the Legendre expansion of the density and the potential. The full parameter space is now four-dimensional, with two additional parameters, defining the shape of the rotation profile. Three rotation regimes can be introduced, namely of moderate, rapid, and extreme rotation. However, significant variations in the structure of the models are primarily associated with concentration and central rotation strength, as for the previous family. We explored the properties of the configurations resulting from two options for the truncation prescription, with emphasis on the family which, in the limit of vanishing internal rotation, reduces to the spherical limit of the models proposed by Wilson (\cite{Wil75}). In particular, configurations in the rapid and extreme rotation regimes exhibit a central toroidal structure, the volume of which increases with the value of the central rotation strength parameter. By making use of asymptotic expansions of the density, mean velocity, and pressure tensor components for small radii, we found the condition for the existence of such central toroidal structure, as well as the condition for the maximum value of the central rotation strength parameter admitted by a configuration with a given concentration. 
\item The differentially rotating models show a variety of realistic velocity dispersion profiles, characterized by the presence of pressure isotropy and radially-biased anisotropy in the central and intermediate regions, respectively. The kinematical behavior in the outer parts depends on the adopted truncation prescription; in particular, the family which, in the nonrotating limit, reduces to the Wilson spheres is characterized by tangentially-biased anisotropy.  This kinematical feature (rarely obtained in equilibrium models) is of great interest for two reasons: (i) Tangentially-biased pressure anisotropy is observed in the presence of internal rotation in globular clusters. For example, the full three-dimensional view of the velocity space of $\omega$ Cen, obtained from proper motions and radial velocities measurements, has revealed that this object is characterized by significant rotation and tangential anisotropy in the outer parts (van de Ven et al. \cite{Ven06}). (ii) The dynamical evolution of a cluster in a tidal field is known to induce a rapid development of tangential anisotropy in the outer parts of the stellar system. In fact, if a cluster fills its Roche lobe and starts losing mass, there is a preferential loss of stars on radial orbits induced by the external tidal field at large radii, where tangential anisotropy in velocity space is thus established (Takahashi \& Lee \cite{TakLee00}; Baumgardt \& Makino \cite{BauMak03}).       
\item The presence of differential rotation may induce nontrivial gradients in the line-of-sight velocity dispersion profile of a stellar system, even if the amount of rotation is modest. Therefore, this important physical ingredient should be taken into account properly. In this respect, dynamical studies of globular clusters and other low-mass stellar systems by means of models based on the use of the Jeans equations are less satisfactory, because, at least in their most popular (nonrotating) application, they are used to reproduce variations in the slope of the kinematical profile of a system only by means of a (sometimes significant) amount of pressure anisotropy. 
\item As expected, differential rotation also induces nontrivial deviations from spherical symmetry; in fact, the models are characterized by a great variety of (projected) ellipticity profiles, dependent on the combined effect of concentration and central rotation strength. Configurations in the moderate rotation regime are characterized by realistic nonmonotonic ellipticity profiles (e.g., see Geyer et al. \cite{Gey83}), while models in the rapid rotation regime have monotonically decreasing profiles. To some extent, this morphological feature is complementary to that of the uniformly rotating models, in which the configurations are always characterized by monotonically increasing ellipticity profiles. Interestingly, the behavior of the ellipticity profiles is not necessarily correlated with the mean line-of-sight velocity profiles, that is configurations with a nonmonotonic mean velocity profile may have a monotonic ellipticity profile.        
In addition, the isophotes of the relevant surface density distribution tend to be characterized by a ``boxy'' structure.        
\item From a comparison of the equilibrium configurations resulting from two options for the truncation prescription of the family of differentially rotating models, we confirm that the interplay between internal rotation, anisotropy in velocity space, and truncation in phase space is highly nontrivial. In fact, as also noted by Hunter (\cite{Hun77}), the structure of the outer parts of a model is particularly sensitive, both from the morphological and the kinematical point of view, to the adopted truncation. One way to select the most appropriate truncation from the physical point of view will be to address the issue in the context of formation and evolution of the class of stellar systems under consideration (see also last item below).
\item Models in the moderate rotation regime seem to be particularly appropriate for describing rotating globular clusters, since the relevant configurations are characterized by a number of realistic properties, such as the presence of nonmonotonic ellipticity profile, the behavior of surface density profile in the outer parts  similar to the one associated with spherical Wilson models, the existence of pressure isotropy in the central regions and tangentially-biased anisotropy at the boundary, as well as realistic values of the ratio $\langle v_P \rangle/\sigma_P$. In a following paper, we plan to apply our family of differentially rotating models to selected Galactic globular clusters that show the presence of significant rotation, such as $\omega$ Cen and 47 Tuc.  
\item Configurations with strong differential rotation, characterized by the presence of a sizable central toroidal structure and by a off-centered peak of the surface brightness profiles, may be useful to shed light on the internal dynamics of the so-called ``ring clusters''. This class of object, originally observed in the Small Magellanic Cloud (Hill \& Zaritsky \cite{HilZar06}) and subsequently noted also in the Large Magellanic Cloud (Werchan \& Zaritsky \cite{WerZar11}), is characterized by a sizable dimple of the central surface brightness, resulting in an off-centered peaked density profile. A proper dynamical interpretation of these objects is currently missing. 
\item The families of models illustrated in the present paper may also help to clarify the role of angular momentum in the formation and dynamical evolution of globular clusters. The results of an extensive survey of N-body simulations, designed to study the dynamical stability and the long term evolution of the models described here, will be presented in follow-up papers (Varri et al. \cite{VarAAS}; Varri et al. in preparation).  

\end{itemize}
\begin{acknowledgements}
We would like to thank E. Vesperini for a critical reading of the manuscript and a number of valuable suggestions, and S.~L.~W. McMillan, D. Heggie, L. Ciotti, J. Anderson, F. Rasio, G. Lodato, A. Sollima, P. Bianchini, and A. Zocchi for interesting conversations. ALV gratefully acknowledges the hospitality of the Department of Physics at Drexel University, where part of the article was written, under the support of a US-Italy Fulbright Visiting Student Researcher Fellowship. This work is partly supported by the Italian MIUR.
\end{acknowledgements}


\appendix

\section{Spherical limit of rotating models}
\label{AppA}

The spherical nonrotating limit of the families of rotating models considered in the present paper are defined by the following distribution functions
\begin{equation}\label{fPTS}
f_{PT}(E)= A\mathrm{e}^{-aE_0}\mathrm{e}^{-a(E-E_0)}~,
\end{equation} 
\begin{equation}\label{fKS}
f_K(E)= A\mathrm{e}^{-aE_0}\left[\mathrm{e}^{-a(E-E_0)}-1\right]~,
\end{equation} 
\begin{equation}\label{fWS}
f_{WT}(E)= A \mathrm{e}^{-aE_0}\left[\mathrm{e}^{-a(E-E_0)} - 1 + a(E-E_0)\right]~,
\end{equation}
if $E \le E_0$ and $f_i(E) = 0$ otherwise (in the following, the index $i=1,2,3$ denotes the models defined by Eqs.~(\ref{fPTS})-(\ref{fWS}), in the same order). In particular, $f_K(E)$ defines the King (\cite{Kin66}) models and represents the spherical nonrotating limit of the family of rigidly rotating models defined by $f_K^r(H)$ (see Eq.~(\ref{fK})); $f_{WT}(E)$ and $f_{PT}(E)$ are the spherical isotropic limit of the Wilson (\cite{Wil75}) and Prendergast \& Tomer (\cite{PreTom70}) models, and represent the spherical nonrotating limit of our differentially rotating models defined by $f_{WT}^d(I)$ (see Eq.~(\ref{fW})) and $f_{PT}^d(I)$ (see Eq.~(\ref{fPT})), respectively.  In the previous expressions, $A$, $a$, are positive constants, defining two dimensional scales, while $E_{0}$ is the cut-off energy, which implies the existence of a truncation radius $r_{tr}$ for the spherical system. For all the families of models considered here one important parameter is the central concentration, as measured by the depth of the dimensionless central potential well $\Psi=\psi(0)=a[E_0-\Phi(0)]$ or by the parameter $c=\log(\hat{r}_{tr})$.  Note that, if we compare three models (one for each family) having the same value of $\Psi$, the corresponding values of $c$ are not the same, that is the relevant dimensionless truncation radii are different ($\hat{r}_{tr}$ increases  from $f_{PT}(E)$, to $f_K(E)$, to $f_{WT}(E)$). In other words, the structure of the outer parts of a spherical isotropic truncated model strictly depends on the truncation prescription; this property is particularly relevant for the interpretation of the photometric profiles of globular clusters (see the systematic comparison between spherical King and Wilson models performed by McLaughlin \& van der Marel \cite{LauMar05}).     

From the integration of the distribution functions in velocity space, the corresponding intrinsic density distributions are recovered. Using the same dimensionless units introduced in the main text, we denote the dimensionless density profiles by $\hat{\rho}_{i,S}=\rho_{i,S}/\hat{A}$ (for the definition of $\hat{A}$, see Eq.~(\ref{hatA})), with     
\begin{equation}\label{rhoS}
\hat{\rho}_{i,S}(\psi)=D_i\,\mathrm{e}^{\psi}\gamma\left(E_i,\psi\right)~,
\end{equation}
here $\psi$ indicates the dimensionless escape energy, defined as in Eq.~(\ref{psidiff})~.

Similarly, the trace of the pressure tensor in dimensionless form (divided by a factor 3) can be written as $\hat{p}_{i,S}={(a/\hat{A})p}_{i,S}$, where
\begin{equation}\label{pS}
\hat{p}_{i,S}(\psi)=P_i\,\mathrm{e}^{\psi}\gamma\left(E_i+1,\psi\right)~;
\end{equation}
$D_i$, $P_i$, and $E_i$ are numerical coefficients resulting from the integration in velocity space and are summarized in Tab.~\ref{SphCoeff}. Note that the coefficient appearing as first argument of the incomplete gamma function in the pressure profile is related to the corresponding coefficient in the density profile, because they are, respectively, the second-order and zeroth-order moment of the distribution function in velocity space.
\begin{table}[h]
    \caption[]{\normalsize Coefficients for the spherical nonrotating models}
       \label{SphCoeff}
       \begin{center}
       \begin{tabular}{lccc}
          \hline
          \hline          
          \noalign{\smallskip}
          Model      &  D & P & E\\
          \noalign{\smallskip}
          \hline
          \noalign{\smallskip}
		Prendergast-Tomer	& 	$3/2$	&	$1$	&	$3/2$\\
		King	&	$1$	&	$2/5$	& 	$5/2$\\
		Wilson	&	$2/5$	&	$4/35$	&	$7/2$\\	
          \noalign{\smallskip}
          \hline
       \end{tabular}
       \end{center}
\end{table}

\section{Plain-truncated rotating models}
\label{AppB}

We summarize here the intrinsic properties of the family of models defined by $f_{PT}^d(I)$ (see Eq.~(\ref{fPT})), for a comparison with the corresponding quantities derived from the family of models defined by $f_{WT}^d(I)$ (see Eq.~(\ref{fW})); a full description is given by Varri (\cite{thesis}). The density profile in dimensionless form can be written as 
\begin{equation}\label{rhoPT}
\hat{\rho}_{PT}(\hat{r},\theta,\psi)=\hat{\rho}_{WT}(\hat{r},\theta,\psi)+\psi^{3/2}+\frac{2}{5}\psi^{5/2}~;
\end{equation}
the asymptotic behavior of the density profile in the outer parts ($\psi \rightarrow 0$), with respect to the dimensionless escape energy, is given by:
\begin{equation}
\hat{\rho}_{PT}(\hat{r},\theta,\psi)=\psi^{3/2}+\frac{2}{5}\psi^{5/2}\left(
1+\frac{9}{2}\chi\hat{r}^2\sin^2{\theta} \right) +\mathcal{O}(\psi^{7/2})~.
\end{equation}
In the central region ($\hat{r}\rightarrow 0$), the expansion to second order in radius is
\begin{eqnarray}
&&\hat{\rho}_{PT}(\hat{r},\theta,\Psi)=\hat{\rho}_{PT,0} + \frac{1}{2}\left[ 
9\chi\hat{r}^2\sin^2{\theta} \mathrm{e}^{\Psi}\gamma\left(\frac{5}{2},
\Psi\right) + \nonumber \right.\\
&& \left. \left. \frac{3}{4}\mathrm{e}^{\Psi}\gamma\left(\frac{1}{2},\Psi\right)
\frac{\partial^2 \psi}{\partial \hat{r}^2}\right|_0 \right]\hat{r}^2 +\mathcal{O}(\hat{r}^4)~,
\end{eqnarray}
where the central value is $\hat{\rho}_{PT,0}=3/2\,\mathrm{e}^{\Psi}\gamma\left(3/2,\Psi\right)=\hat{\rho}_{PT,S}(\Psi)$, consistent with the value of the corresponding nonrotating model. 

The mean velocity is in the azimuthal direction. In dimensionless units it is given by 
\begin{equation}\label{uPT}
{\langle\hat{v}_{\phi}\rangle}_{PT}(\hat{r},\theta,\psi)=\frac{3\,\mathrm{e}^{\psi}}{2^{3/2}\,\hat{\rho}_{PT}} \int_0^{\psi} ds\,
\mathrm{e}^{-s} s \int_{-1}^{+1}dt\,t\,g(s,t,\hat{r},\theta)~,
\end{equation}
which, in the outer parts of the models reduces to
\begin{equation}\label{uPTbou}
{\langle\hat{v}_{\phi}\rangle}_{PT}(\hat{r},\theta,\psi)=\psi\frac{6}{5}{\chi}^{1/2}\hat{r}\sin{\theta}+\mathcal{O}(\psi^{3/2})~.
\end{equation}
The expansion for $\hat{r}\rightarrow 0$ to first order in radius can be written as
\begin{equation}
{\langle\hat{v}_{\phi}\rangle}_{PT}(\hat{r},\theta,\Psi)=2\frac{\gamma(5/2,\Psi)}{\gamma(3/2,\Psi)}
\chi^{1/2}\hat{r}\sin{\theta}+\mathcal{O}(\hat{r}^3)~;
\end{equation}
at variance with the family defined by $f_{WT}^d(I)$, the dimensionless angular velocity depends also on the concentration parameter.

As far as the pressure tensor is concerned, we recall that by construction $\hat{p}_{rr}=\hat{p}_{\theta \theta}$. The dimensionless radial component is given by
\begin{equation}\label{prrPT}
\hat{p}_{PT,rr}(\hat{r},\theta,\psi)=\hat{p}_{W,rr}(\hat{r},\theta,\psi)+\frac{2}{5}\psi^{5/2}+\frac{4}{35}\psi^{7/2}~,
\end{equation}
which, expanded in $\psi$ (as is appropriate for the outer parts), reduces to
\begin{eqnarray}
&&\hat{p}_{PT,rr}(\hat{r},\theta,\psi)=\frac{2}{5}\psi^{5/2}+\nonumber\\
&&\frac{4}{35}\psi^{7/2}\left(1+ \frac{9}{2}\chi\hat{r}^2\sin^2{\theta} \right) +\mathcal{O}(\psi^{9/2})~.
\end{eqnarray}
In the inner parts it can be approximated to second order in radius, by the following expression
\begin{eqnarray}
&&\hat{p}_{PT,rr}(\hat{r},\theta,\Psi)=\hat{p}_{PT,0} + \frac{1}{2}\left[ 
\frac{18}{5}\chi\hat{r}^2\sin^2{\theta} \mathrm{e}^{\Psi}\gamma\left(\frac{7}{2},\Psi\right) \right. \nonumber \\
&& \left. \left.+ \frac{3}{2}\mathrm{e}^{\Psi}\gamma\left(\frac{3}{2},\Psi\right)
\frac{\partial^2 \psi}{\partial \hat{r}^2}\right|_0\right]\hat{r}^2 +\mathcal{O}(\hat{r}^4)~,
\end{eqnarray}
where the central value is given by $\hat{p}_{PT,0}=\mathrm{e}^{\Psi}\gamma(5/2,\Psi)=\hat{p}_{PT,S}(\Psi)$, consistent with the value of the corresponding model in the limit of vanishing rotation.

The azimuthal component of the pressure tensor can be written as 
\begin{eqnarray}\label{pffPT}
\hat{p}_{PT,\phi\phi}(\hat{r},\theta,\psi)&=&\hat{p}_{W,\phi\phi}(\hat{r},\theta,\psi)+\frac{2}{5}\psi^{5/2}+\frac{4}{35}\psi^{7/2}+\nonumber\\
&&\hat{\rho}_{WT}{\langle\hat{v}_{\phi}\rangle}_{WT}^2-\hat{\rho}_{PT}{\langle\hat{v}_{\phi}\rangle}_{PT}^2~,
\end{eqnarray}
which at the boundary is approximated by
\begin{eqnarray}
&&\hat{p}_{PT,\phi\phi}(\hat{r},\theta,\psi)=\frac{2}{5}\psi^{5/2}+ \nonumber\\
&&\frac{4}{35}\psi^{7/2}
\left(1- \frac{9}{10}\chi\hat{r}^2\sin^2{\theta} \right)+\mathcal{O}(\psi^{9/2})~.
\end{eqnarray}
Close to the center we find 
\begin{eqnarray}
&&\hat{p}_{PT,\phi\phi}(\hat{r},\theta,\Psi)= \hat{p}_{PT,0} + \frac{1}{2}
\left\{ 6{\chi}\sin^2{\theta}\left[ \frac{9}{5}\mathrm{e}^{\Psi}
\gamma\left(\frac{7}{2},\Psi\right)-\right. \right.\nonumber \\
&&\left. \left. \left. 2\mathrm{e}^{\Psi}\frac{\gamma(5/2,\Psi)^2}{\gamma(3/2,\Psi)} \right] 
+ \frac{3}{2}\mathrm{e}^{\Psi}\gamma\left(\frac{3}{2},\Psi\right)
\frac{\partial^2 \psi}{\partial \hat{r}^2}\right|_0\right\}\hat{r}^2 +\mathcal{O}(\hat{r}^4)~.
\end{eqnarray}

Therefore, the models in this family are characterized by pressure isotropy in the central region, radially-biased pressure anisotropy in the intermediate part, and pressure isotropy at the boundary (since, for both $\hat{p}_{rr}$ and $\hat{p}_{\phi\phi}$, the term of lower order in $\psi$ is given by $2/5\,\psi^{5/2}$), at variance with the family defined by $f_{WT}^d(I)$ in which tangentially-biased anisotropy is present. 

Using the asymptotic expression of the density in the central regions recorded above, in this case, the condition for the existence of the central toroidal structure is given by   
\begin{equation}
\chi > \frac{\gamma(1/2,\Psi)}{12\gamma(5/2,\Psi)}\left[3 +\frac{C_2}{2}\left(\frac{5}{2}\right)^{1/2} \right]~,
\end{equation}
where $C_2$ is defined as in Eq.~(\ref{C2}). By evaluating the sign of the velocity and pressure term in the radial component of the Jeans equation (see Eq.~(\ref{jeans})), we found that, in this case, the requirement of positivity of the velocity term is just a necessary but {\em not} sufficient condition for the existence of the central toroidal structure. In other words, in this family, configurations with angular velocity higher than the angular velocity associated with the circular orbit of a single star but {\em without} a central toroidal structure can exist. The condition for maximally rotating configurations, which, in this case, is given by
\begin{equation}
\chi< \frac{\gamma(1/2,\Psi)}{8\,\gamma(5/2,\Psi)}\left(\frac{5}{2}\right)^{1/2}|C_2|~,
\end{equation} 
completes the summary of the intrinsic properties of the family of plain-truncated differentially rotating models.

\section{The numerical procedure for the construction of differentially rotating models}
\label{AppC} 

The construction of the equilibrium configurations for the families of differentially rotating models, defined by Eqs.~(\ref{fW}) and (\ref{fPT}), has been performed by numerically solving the relevant Poisson equation as a nonlinear equation for the unknown potential $\psi$. The code which implements the iteration procedure described in Sect.~\ref{Diff2} starts with the calculation of the ``seed solution'', given by a selected spherical configuration from the family of models that represent the limit in the case of vanishing internal rotation of the family of interest (i.e., defined by Eqs.~(\ref{fWS}) and (\ref{fPTS}), respectively). Such spherical solution is used, in the first step of the iteration, to evaluate  (by means of a Double Gaussian Quadrature) the density distribution, given by $\hat{\rho}_{WT}$ or $\hat{\rho}_{PT}$, on a spherical grid in the meridional plane, defined by $(r_i,\theta_j)$. The grid is linear in the radial coordinate and the angular positions are defined by
\begin{equation}
\theta_j=\frac{\pi(2j+1)}{4\,l_{max}}~,
\end{equation}     
where $j=1,...,2l_{max}+1$. Typically, we used $\approx 300$ radial steps and we set $l_{max}=21$, in order to have sufficient accuracy to describe the complex morphologies of the configurations in the rapid and extreme rotation regimes. The discrete direct and inverse Legendre transforms, required at each step of the iteration for the calculation of the density and potential coefficients, defined by Eqs.~(\ref{rhon})-(\ref{psin}), are performed (up to the order $l_{max}$) by means of a package based on S2kit 1.0 by Kostelec \& Rockmore (\cite{KosRoc04}), which makes use of FFTW 3.2.1 by Frigo \& Johnston (\cite{FriJoh05}). The Cauchy problems for the potential radial functions expressed in Eqs.~(\ref{eqint0})-(\ref{eqintn}) are evaluated by means of a numerical integration with Romberg's rule. The convergence condition for the solution at the $n$-th step of the iteration is formally defined as
\begin{equation}
\frac{\psi_{ij}^{(n)}-\psi_{ij}^{(n-1)}}{\psi_{ij}^{(n)}}< 10^{-3}~,
\end{equation}      
for every $i,j$, where $\psi_{ij}^{(n)}=\psi^{(n)}(r_i,\theta_j)$; about 10 (25) iteration steps are needed for the construction of configurations characterized by low (high) values of $\chi$. The accuracy of the solutions found with our code has been checked by the following tests: (i) the virial theorem is satisfied with accuracy of the order of $10^{-4}$ or better; (2) the radial component of the Jeans equation is satisfied with the accuracy of the order of $10^{-3}$ or better; (3) the asymptotic behaviors, both in the central and in the outer parts, of all the moments of the distribution function are confirmed.    

\end{document}